\documentclass{JHEP3}
\usepackage{graphicx}
\usepackage{ifthen}
\newcommand{\newc}{\newcommand}
\def\be{\begin{equation}}
\def\ee{\end{equation}}
\def\bea{\begin{eqnarray}}
\def\ena{\end{eqnarray}}
\def\bedm{\begin{displaymath}}
\def\endm{\end{displaymath}}
\def\cah{{\cal H}}

\newc{\blm}[1]{\mbox{\boldmath$#1$}}
\newc{\ERR}[3]{Erratum-ibid.\ {\bf #1}, (#3) #2}
\def\vfu{V_{\!\mathit{full}}}
\newc{\roa}[1]{\roarrow{#1}}
\newc{\arghq}{(H_1^0, H_2^0;\blm{0})}
\newc{\fvevq}{(\lg H_1^0 \rg, \lg H_2^0 \rg; \blm{\chi})}
\newc{\argf}{(H_1^0, H_2^0,\blm{\chi}; Q)}
\newc{\argh}{(H_1^0, H_2^0, \blm{0}; Q)}
\newc{\fvev}{(\lg H_1^0 \rg, \lg H_2^0 \rg, \blm{\chi}; Q)}
\newc{\hvev}{(\lg H_1^0 \rg, \lg H_2^0 \rg; \blm{0})}
\newc{\magf}[1]{| #1 |^2}
\def\lg{\langle}
\def\rg{\rangle}
\def\lpa{\left(}
\def\rpa{\right)}
\def\lbs{\left|}
\def\rbs{\right|}
\def\cH{H^*}
\newc{\h}[1]{\ifthenelse{\equal{#1}{1}}{h_1}{h_2}}
\newc{\ch}[1]{\ifthenelse{\equal{#1}{1}}{h^*_1}{h^*_2}}
\def\sEC{\tilde{E}^c}
\def\csEC{\tilde{E}^{c*}}
\def\sLE{\tilde{L}}

\def\sle{\tilde{\ell}}
\def\csle{\tilde{\ell}^*}
\def\sne{\tilde{\nu}}
\def\csne{\tilde{\nu}^*}
\def\sUC{\tilde{U}^c}
\def\csUC{\tilde{U}^{c*}}
\def\sDC{\tilde{D}^c}
\def\csDC{\tilde{D}^{c*}}
\def\sQU{\tilde{Q}}

\def\squ{\tilde{u}}

\def\sqd{\tilde{d}}

\def\lmv{\mathcal{L}_V}
\def\msv{\mathcal{M}_V^2}
\newc{\snw}[1]{\ifthenelse{#1 > 1}{\sin^{#1}\theta_W}{\sin\theta_W}}
\newc{\csw}[1]{\ifthenelse{#1 > 1}{\cos^{#1}\theta_W}{\cos\theta_W}}
\newc{\eg}{{\it e.g.}\/}
\newc{\ie}{{\it i.e.}\/}
\newc{\etal}{{\it et.al.}\/}
\title{On Nonstandard Vacuua in Minimal Supergravity Models.}
\author{D.V.~Gioutsos and C.E.~Vayonakis \\
        Division of Theoretical Physics,University of Ioannina \\
        Ioannina, GR - 451 10, GREECE \\
   Email: \email{dgioutso@cc.uoi.gr}, \email{cevayona@cc.uoi.gr}}
\abstract{The plethora of scalar fields participating in the formulation of a
softly broken supersymmetric theory can threat the stability of the standard
vacuum. The generic situation is twofold. Directions in scalar field space may
exist along which the potential becomes unbounded from below or local minima
deeper than the standard one are likely to appear where charge and/or color
symmetries are broken. An investigation of these matters requires a thorough
study of the tree level effective potential as well as its radiative
corrections. However, the presence of many different mass scales in such a
generic supersymmetric theory needs an appropriate renormalization improved
treatment. Combining the decoupling theorem with a conveniently chosen
renormalization scale for each field configuration, the well known multimass
scale problem is circumvented. In this context, the ordinary universal
soft-parameters case of the minimal supersymmetric model as well as a
non-universal case of a brane world minimal supergravity model are examined
closely for unphysical vacuua.}
 \preprint{\hepph{0208028}}
\begin{document}
\section{\label{Intro}Introduction}

Supersymmetric (SUSY) theories are the best motivated extensions of the
Standard Model (SM) of the electroweak and strong interactions. They provide an
elegant way to stabilize the huge hierarchy between the Grand Unification (GUT)
and the Fermi scale, and predict a variety of new matter states (sparticles)
creating a natural framework to cancel the quadratic divergences of the
radiative corrections to the Higgs boson mass. One can classify these models by
the mechanism for communicating SUSY breaking from a hidden sector to the
observable sector. Possibilities include gravity mediated SUSY breaking (SUGRA)
\cite{nilles}, gauge mediated SUSY breaking, \cite{gaugemed} and anomaly
mediated SUSY breaking \cite{anomaly}.

The so-called minimal supergravity (mSUGRA) model  has traditionally been the
most popular choice for phenomenological SUSY analyses. In mSUGRA, it is
assumed that its most economical low energy realization, the so called Minimal
Supersymmetric Standard Model (MSSM) \cite{mssm}, is valid from the weak scale
all the way up to the GUT scale $M_X\simeq 2\times 10^{16}$ GeV, where this
choice is usually taken due to apparent gauge coupling unification. In the MSSM
one assumes a minimal gauge group, \ie~${\rm SU(3)_C}\times {\rm SU(2)_L}
\times {\rm U(1)_Y}$, and a minimal particle content, \ie~three generations of
fermions (no right handed neutrinos included) and their spin zero partners as
well as two Higgs doublet superfields to break the electroweak symmetry. In
order to explicitly break supersymmetry while preventing the reappearance of
quadratic divergences, a collection of soft terms \cite{soft} is added to the
Lagrangian: mass terms for the gauginos ($m_{1/2}$), mass terms for the scalar
fermions ($m_0$), mass and bilinear terms for the Higgs bosons ($B_0$), and
trilinear couplings ($A_0$) between sfermions and Higgs bosons. In the general
case, that is if one allows for intergenerational mixing and complex phases,
the soft SUSY breaking terms will introduce a huge number of unknown
parameters. This feature makes any phenomenological analysis in the general
MSSM a daunting task. In this context, severe phenomenological constraints are
imposed on the parameter space by flavor changing neutral currents (FCNC) and
CP violation \cite{fc-cp} as well as unphysical vacuua.

To reduce the number of free parameters, one needs an explanation of how
supersymmetry is broken. In mSUGRA, supersymmetry is spontaneously broken via a
hidden sector field vacuum expectation value (VEV), and the SUSY breaking is
communicated to the visible sector via gravitational interactions. For a flat
K\"ahler metric $G_i^j$ and common gauge kinetic functions $f_{AB}$, this leads
to universal values for $m_0$, $m_{1/2}$ and $A_0$, $B_0$ at the GUT scale
$M_{X}$. This assumption of universality in the scalar sector leads to the
phenomenologically required suppression of flavor violating processes that are
supersymmetric in origin. However, there is no known physical principle which
gives rise to the desired form of $G_i^j$ and $f_{AB}$; indeed, for general
forms of $G_i^j$ and $f_{AB}$, non-universal masses are expected. Hence, the
universality assumption is regarded as being entirely motivated by the
phenomenological need for suppression of flavor violating processes in the
MSSM.

In recent years the branes, which are typical in models with extra dimensions
\cite{brani}, have been found to fit naturally with the idea of breaking
supersymmetry via a hidden sector. The basic five-dimensional setup of brane
world models \cite{brmo} is that of four-dimensional hypersurfaces (branes)
hosting familiar gauge and matter fields, which are embedded in a higher
dimensional ambient space, the bulk, populated by gravitational and
gauge-neutral fields. The bulk degrees of freedom couple to the fields living
on branes through various types of interactions, but the effects of
SUSY-breaking will be communicated gravitationally to the observable world
through the five-dimensional interior bulk. This scenario thereby places the
question of SUSY-breaking in an entirely new geometric context.

Brane world SUSY breaking generally gives rise to tree-level soft scalar masses
for visible sector squark and slepton fields. These masses are generally not
universal. Without additional assumptions about flavor, these non-universal
scalar mass matrices are not necessarily aligned with the quark and lepton mass
matrices, and dangerous sflavor violation can occur. A model to overcome this
flavor problem has been recently proposed. The authors of Ref.~\cite{shafi}
have suggested a gravity mediated SUSY breaking in 5D spacetime with two 4D
branes B1 and B2 separated in the extra dimension. The SUSY breaking effects
from the hidden sectors localized on the branes are transmitted through
gravitational couplings both to the visible sector fields located on the same
brane and to bulk gauginos. Thus, if the SUSY breaking scales are $m_1$ and
$m_2$ at two separated branes B1 and B2 respectively, the localized chiral
multiplets at B1 (B2) could get soft SUSY breaking effects of order $m_1$
($m_2$). Assuming that the SUSY breaking scales at B1 and B2 are $m_1$ ($\sim$
10--20 TeV) $>$ $m_2$ ($\sim$ 1 TeV), the superpartners of the first two
generations can be easily made sufficiently heavy to overcome the SUSY flavor
problem by locating them on B1. In order to keep the radiative corrections to
the Higgs masses under control, the third family must reside on B2. The two
Higgs multiplets should reside in the bulk, so that the first two and the third
generations of the quarks and leptons can couple to them at B1 and B2,
respectively.

Beyond the SUSY flavor requirements, an additional constraint on mSUGRA
parameters can be obtained by demanding that the global minimum of the scalar
potential is indeed the minimum that leads to appropriate electroweak symmetry
breaking \cite{rad}. From a theoretical point of view, the plethora of new
scalar fields which are introduced in MSSM may lead to many possible directions
in field space where field configurations could be developed deeper than the
standard minimum. The generic situation
\cite{sher,casas,vacuua1,vacuua2,vacuua3} is that the scalar potential can
receive negative contributions from quadratic or cubic term in the
supersymmetry breaking Lagrangian if the usually dominated quartic $D$ and $F$
terms are suppressed. A systematic classification of all possible dangerous
directions in the scalar field space that can potentially lead to undesirable
minima has been done by the authors of Ref.~\cite{casas}. These directions have
been categorized either as field directions that are unbounded from below (UFB)
or as directions that lead to charge or color breaking minima (CCB).

The purpose of this work is to extend and complement the work done on this
subject by analyzing the impacts the one loop radiative corrections have on
these unphysical configurations for the models described above. Yet, the
effective potential in which the vacuum structure is encoded, is a poorly known
object beyond the tree-level approximation. One reason for this is the
dependence of its loop corrections upon the very many different mass scales
present in these models, so that a renormalization group analysis becomes
rather tricky. In general, when one deals with a system possessing a large mass
scale $Q_M$, compared with the scale $Q_{\mu}$ at which one discusses physics,
large logarithms such as $\ln(Q_M/Q_{\mu})$ always appear which affect the
convergence of the loop expansion. In this situation, one considers resuming
the perturbation series by using the renormalization group equation (RGE).
Nonetheless, in many realistic applications one often has to deal with an
additional mass scale $Q_m$ with the hierarchy $Q_{\mu} \ll Q_m \ll Q_M$. For
instance, one can regard $Q_{\mu}, Q_m, Q_M$ as the weak,
supersymmetry-breaking and unification scales respectively. When we study such
a system, we face the problem of multimass scales \cite{multisc}: There appear
several types of logarithms $\ln(Q_M/Q_{\mu})$ and $\ln(Q_m/Q_{\mu})$, while we
are able to sum up just a single logarithm by means of the RGE.

In the present work, trying to circumvent this problem, a generalized improving
procedure based on Refs~\cite{quir,dvg} is applied to the relevant effective
potential. The main idea of the method is to make use of the decoupling theorem
\cite{caraz} combined with a conveniently chosen renormalization scale $Q^*$
for each field configuration. By this procedure, it is made possible to treat
essentially a single log factor at any renormalization scale, since all the
heavy particles (heavier than that scale) decouple and all the light particles
(lighter than that scale) yield more or less the same log factors.

The rest of this paper is organized as follows. After setting our notation and
conventions in Sec.~\ref{setup}, an estimation of the physical vacuum is
attempted in various effective low energy SUSY models in Sec.~\ref{asp}. Next,
section \ref{ddnut} describes briefly the most dangerous directions that could
lead to unphysical situations and Sec.~\ref{calsch} is an outline of our
calculational scheme. In Sec.~\ref{resl} we give a brief summary of our results
and section \ref{conc} summarizes our conclusions. Finally, detailed formulae
for the various mass matrix elements involved in one loop expressions of the
effective potential are presented in Appendices.
\section{\label{setup}The Physical Set-up}

Let us briefly review some of the basic ingredients required for our analysis.
A globally supersymmetric and $SU(3)_C\otimes SU(2)_L\otimes U(1)_Y$ gauge
invariant Lagrangian with minimal content can be constructed from the usual
$R$-symmetry conserving superpotential\footnote
{$i,j$ are $SU(2)$ indices, $a$ is a color index (family indices are
suppressed). Also $\epsilon_{12} = +1$.}
 \be
 \label{super}
 \mathcal{W}= \blm{Y_e} {\hat L}^j {\hat E}^c {\hat\cah}_1^i
 \epsilon_{ij}
         + \blm{Y_d} {\hat Q}^{ja} {\hat D}_a^c {\hat\cah}_1^i \epsilon_{ij}
         + \blm{Y_u} {\hat Q}^{ja} {\hat U}_a^c {\hat\cah}_2^i \epsilon_{ij}
         + \mu {\hat\cah}_1^i {\hat\cah}_2^j \epsilon_{ij}
\ee%
where ${\hat Q}^T=(\hat{u}~~\hat{d})$, ${\hat U}^c$, ${\hat D}^c$ are the Quark
Superfields, ${\hat L}^T=(\hat{\nu}~~\hat{\ell})$, ${\hat E}^c$ are the Lepton
Superfields and ${\hat\cah}^T_1=(\hat{H}_1~~\hat{h}_1)$,
${\hat\cah}^T_2=(\hat{h}_2~~\hat{H}_2)$ are the Higgs Superfields. Generally
the Yukawa matrices $\blm{Y_u}$, $\blm{Y_d}$, $\blm{Y_e}$ and the parameter
$\mu$ are complex.

In order to explicitly break supersymmetry as required by experiment, while
keeping quadratic divergences suppressed, a collection of soft terms is added
to the Lagrangian. These include: mass terms for all scalar fermions and
gauginos, bilinear terms for the Higgs bosons and trilinear interactions
between sfermions and Higgses. Consequently, the most general soft SUSY
breaking Lagrangian with real mass terms in this minimal scheme is\footnote
{$\tilde{B}$, $\tilde{W}$, $\tilde{G}$ stand for gauginos (Weyl spinors) and
$\Gamma$, $R$ are $SU(2)$, $SU(3)$ group indices respectively.}
\bea \mathcal{L}_{SOFT}
        &=& -\frac{1}{2} M_1 (\tilde{B} \tilde{B})
            -\frac{1}{2} M_2 (\tilde{W^{\Gamma}} \tilde{W^{\Gamma}})
            -\frac{1}{2} M_3 (\tilde{G^R} \tilde{G^R}) \nonumber \\
        & & -  m_{H_1}^2 |\cah_1|^2
            -m_{H_2}^2 |\cah_2|^2 -m^2_{\tilde{Q}}  |\tilde{Q}|^2
            -m^2_{\tilde{D}^c}  |\tilde{D}^c|^2
            - m^2_{\tilde{U}^c}  |\tilde{U}^c|^2  \nonumber\\
        & & - m^2_{\tilde{L}} |\tilde{L}|^2
            -m^2_{\tilde{E}^c}  |\tilde{E}^c|^2
            -(\blm{h_e}  \tilde{L}^j \tilde{E}^c \cah_1^i \epsilon_{ij}
            + \blm{h_d}  \tilde{Q}^{ja}   \tilde{D}_a^c \cah_1^i \epsilon_{ij}
            \nonumber\\
        & & + \blm{h_u}  \tilde{Q}^{ja} \tilde{U}_a^c \cah_2^i \epsilon_{ij}
               + \mbox{H.c.})
            -(B \mu \cah_1^i \cah_2^j \epsilon_{ij}  + \mbox{H.c.})
\label{softL}
\ena%
Here $\cah_1$, $\cah_2$ are the ordinary Higgs boson doublets, $\tilde{Q}$,
$\tilde{D}^c$, $\tilde{U}^c$, $\tilde{L}$, $\tilde{E}^c$ are the squark-slepton
scalar fields and $\blm{h} \equiv \blm{Y} \blm{A}$, where \blm{A} is a $3
\times 3$ matrix containing the ``soft'' trilinear scalar couplings. All extra
soft parameters except masses are generally complex.

Altogether one would then need more than 100 real parameters to describe SUSY
breaking in full generality. Clearly, some simplifying assumptions are
necessary if we want to make comprehensive scans in parameter space.
Specifically, we shall work with real and diagonal Yukawa couplings where all
non-zero entries are positive. Also $\mu$, $B$ and all trilinear soft couplings
$\blm{A}$ are assumed to be real. In our analysis we shall also keep Yukawas
and trilinear soft couplings from light families, since their contributions to
the one-loop effective potential (our main objective) are not always negligible
for an arbitrary field configuration. For clarity reasons, we simplify the
notation using $Y_u^1 = Y_u,~Y_u^3 = Y_t,~Y_e^1 = Y_e,~Y_d^3 = Y_b$ and
similarly for the trilinear couplings.

A dramatic decrease of the unknown parameters arises when the minimal model
just presented is embedded into various supergravity scenarios. The simplest
case of the MSSM results from coupling with $N=1$ Supergravity. This leads to
the following ``universal'' scenario at a very large scale $M_X \simeq 2 \times
10^{16}$ GeV with: $(1)$ Common gaugino mass : $m_{1/2}$, $(2)$ Common scalar
mass : $m_0$, $(3)$ Common trilinear scalar coupling : $A_0$. This reduces the
number of free parameters describing SUSY breaking to just four : The gaugino
mass $m_{1/2}$, the scalar mass $m_0$, the trilinear and bilinear soft breaking
parameters $A_0$ and $B$, which is conventionally trade with the Higgs VEV
ratio $\tan\beta = \lg H_2^0 \rg / \lg H_1^0 \rg$. We also assume unification
of the gauge couplings at the scale $M_X $, while no specific relation is
assumed for the Yukawa couplings there.

However, more complicated alternatives also exist which give definite
predictions for scalar, gaugino masses and soft trilinear terms appearing in
Eq.~(\ref{softL}). As we have mentioned in the introduction, one such
possibility is the recently proposed 5D supergravity model of
Ref.~\cite{shafi}. In this model the chiral matter content resides on two 4D
branes separated in the extra dimension. Higgs bosons as well as gravity and
gauge multiplets reside in the bulk. It is then shown that the SUSY breaking
effects from the hidden sectors localized on the branes are transmitted to the
visible sector fields located on the same brane and bulk gauginos through
gravitational couplings. Trying to overcome the SUSY flavor problem, two SUSY
breaking scales $m_1, m_2$ are introduced at the separated branes $B_1$ and
$B_2$, respectively. Then the localized chiral multiplets at $B_1(B_2)$ could
get soft SUSY breaking effects of order $m_1(m_2)$. Assuming that the first two
generations reside at $B_1$ and the third one is at $B_2$, the following non
universal soft terms are derived
 \bea
m_{0,i}^2 &=& \frac{8m_1^2}{9} (\delta_{i1}+\delta_{i2}) + \frac{8m_2^2}{9}
\delta_{i3}   \\
A_{0,i}  &=& \frac{4m_1}{3} (\delta_{i1}+\delta_{i2}) + \frac{4m_2}{3}
\delta_{i3}
 \ena
where $i$ runs on three generations. Mass terms for the gauginos can be
generated at tree level in a $D=5$ off-shell SUGRA formalism \cite{zuk}. If
SUSY is spontaneously broken at brane $B_1$, heavy gaugino masses given by
$m_{1/2} \simeq m_1\, (M_*/M_P)$, where $M_*$ is a 5D SUGRA fundamental scale
and $M_P$ is the reduced 4D Planck scale, could give rise to large radiative
corrections to the Higgs mass and spoil the naturalness solution. However, a
not so ``large'' extra dimension, for example of the order of $L^{-1} \sim
10^{16}$ GeV (or $M_* \simeq 10 L^{-1} \sim 10^{17}$ GeV) suppresses the
gaugino masses to 1 TeV scale. Note that the cutoff scale $\Lambda_c$
introduced to obtain an effective 4D theory from the 5D one, is of the order of
the extra dimension. Hence, the effective theory below $\Lambda_c \sim M_X$
resembles the familiar gravity mediated scenario of MSSM.

In order to discuss the physical implications of these models at low-energies,
we have to determine the evolution of the various ``couplings'' at
experimentally accessible energies. For that reason, we use two loop RGEs with
all thresholds of the various particles included \cite{rge2,sak}.
\section{\label{asp}Estimating the physical vacuum at low energies}
Regardless of the physical scenarios describing physics at high energies, low
energy world could be best described by a softly broken supersymmetric theory.
In such a framework, the elegant ideas of supersymmetry, unification and
radiative symmetry breaking coexist. The minimal supersymmetric extension of
standard model (MSSM) incorporates all the above. Because of its minimal
content and the radiatively induced symmetry breaking, it is the most
predictive of analogous theories. Let us denote by $\blm{\phi}$ the scalar
field collection of MSSM. Since MSSM is expected to be a stable theory, its
\textit{full} effective potential $\vfu$ (scale independent) must have a global
minimum (physical vacuum). At the physical vacuum, the only scalar fields that
should acquire non-trivial expectation values (VEV) would be the neutral Higgs
components. Simply speaking, the VEV configuration there should read:\footnote{
Since $\vfu$ is $SU(2)$ symmetric, we can choose either $\lg H_1^0 \rg$ or $\lg
H_2^0 \rg$ to be a real non-trivial number. Then the reality of the other
should arise from the minimization conditions.}
$ \lpa \lg H_1^0 \rg,~\lg H_2^0 \rg;~\blm{0} \rpa$.%

We know that MSSM depends on a minimal set of free parameters ($A_0$, $m_0$,
$m_{1/2}$, $\tan\beta$). On general grounds one expects that, for some values
of them, $\vfu$ might be unbounded from below (UFB) (no physical vacuum). For
others, a global minimum may exist where additional scalar fields acquire
non-trivial VEVs breaking in this way at least one of the other gauge
symmetries $(U(1),SU(3))$, a fact that clear contradicts experiment. How should
we protect the theory from cases like that? The most straightforward way is
direct minimization of $\vfu$ for each ``point'' $(A_0, m_0, m_{1/2},
\tan\beta)$ and exclusion of the cases that give non-physical results.

Let us employ this procedure for a physically stable ``point'' where only
$SU(2)$ is broken. To compute the VEVs we group the scalar fields into two
subsets: $\blm{\phi}= \{H_i^0;\chi_k \}$ where $i=1,2$ and $k$ runs on the rest
scalars. Applying the stationary conditions we get:
\begin{equation}\label{stcs}
  \left. \frac{\partial\vfu\arghq}{ \partial H_i^0}\right|_{(\lg H_1^0 \rg,
  \lg H_2^0 \rg)} =0~~\mathrm{and}~~ \left. \frac{\partial\vfu\fvevq}
  {\partial \chi_{k}} \right|_{\blm{\chi}=\blm{0}} = 0
\end{equation}
From a more practical point of view, one must construct an approximation of
these abstract conjectures using loop expansion \cite{ilio}. Indeed, at
one-loop level the effective potential \cite{coleman} becomes scale dependent
and we can write
\begin{equation}\label{1st}
    \vfu \approx V_{1-\mathit{loop}} \argf \equiv
     \Omega^{'} + V^{(0)} + V^{(1)}
\end{equation}
where $Q$ is the renormalization scale and $\Omega^{'}$ is a field independent
quantity \cite{omega} that should be added to preserve scale invariance at
one-loop level. On the other hand, $V^{(0)}$ and $V^{(1)}$ are the classical
and one-loop corrections to the scalar potential respectively.

Since $V^{(0)}$ is at least a second degree polynomial in $\chi_k$, its first
derivatives with respect to  $\chi_k$ vanish at physical vacuum and the 1-loop
approximation to Eq.~(\ref{stcs}) will become
\begin{equation}\label{1lss}
  \left. \frac{\partial V_{1-\mathit{loop}}\argh}{ \partial H_i^0}\right|_{
  (\lg H_1^0 \rg, \lg H_2^0 \rg)} \approx 0~~\mathrm{and}~~ \left.
  \frac{\partial V^{(1)}\fvev}{\partial \chi_k}\right|_{\blm{\chi}=\blm{0}}
  \approx 0
\end{equation}
In order to proceed further, 
we consider the ``reduced'' function $V_{1-\mathit{loop}}\argh$ and expand
around the exact one-loop solution $(v_1,v_2)$, $\lg H_1^0 \rg = \upsilon_1 +
\delta\upsilon_1$, $\lg H_2^0 \rg = \upsilon_2 + \delta\upsilon_2$ where
$\delta\upsilon_i$ are of higher-loop order. We get
\begin{equation}\label{reduce}
   \left. \frac{\partial V_{1-\mathit{loop}}\argh}{ \partial
   H_i^0}\right|_{(\upsilon_1, \upsilon_2)} = 0
\end{equation}
Solving this system of equations, we get the one-loop approximation of the so
called \emph{physical} vacuum. As an example, we will present the well known
expressions for the physical vacuum at tree level. The associated ``reduced''
function is
\be%
 V^{(0)}\argh = m_1^2 |H_1^0|^2 + m_2^2 |H_2^0|^2
   + 2 m_3^2 (H_1^0 H_2^0) + \frac{g^2+g_2^2}{8}
    (|H_1^0|^2 - |H_2^0|^2)^2
\ee%
and its value at the minimum is \cite{vacuua1}
\be%
V_{\mathit{ph}}^{(0)} = -\frac{\left[ m_1^2 - m_2^2 +(m_1^2 + m_2^2) \cos
2\beta \right]^2}{2(g^2+g_2^2)} \label{physv}
\ee%
where $\tan\beta$ stands for the VEV's ratio at tree level.

We stress here that the ``reduced'' function $V_{1-\mathit{loop}}\argh$
appearing in (\ref{reduce}) can be used solely as far as the computation of
$\upsilon_1,\upsilon_2$ is concerned. It is completely unsuitable for drawing
any conclusions about the kind of $\vfu$'s stationary points. Nevertheless,
these ``reduced'' functions can help us to study more general cases.
Practically, one selects a ``direction'' in the field space\footnote{%
Of course, now all the scalar fields are allowed to participate}%
~whose ``reduced'' function collects the deepest values of $\vfu$, at the
approximation level we work, and excludes its trace from the model's parameter
space whenever it leads to a non-physical picture.

\section{\label{ddnut}Classification of dangerous directions in a nutshell}

Scalar potential in SUSY models receives contributions from three sources:
$D$-terms, $F$-terms and soft-breaking terms. $D$-terms provide a quartic
contribution $V\sim\lambda\phi^4$ with $\lambda\geq 0$. The special case
$\lambda=0$ can only occur along special directions in field space known as
``$D$-flat directions''. Along such a flat direction the quartic potential
takes the form $V\sim(\phi_1^2- \phi_2^2)^2\to 0$ as $|\phi_1|\to|\phi_2|$,
which may render $|\phi_i|$ running off to infinity. Whether or not this
occurs, will depend on the magnitude of the other contributions ($F$-terms,
soft terms).

$F$-terms, in turn, due to their supersymmetric nature contribute quadratic,
cubic, and quartic terms to the potential. This does not mean, however, that
the minimum of the potential lies at the origin. Rather, it means that
directions in field space with non-zero quartic $F$ contributions will be
well-behaved far away from the origin. Still, there will be a subset of the
$D$-flat directions which are also $F$-flat and whose behavior will be
completely controlled by soft-breaking terms.

The soft mass contributions, on the other hand, are problematic, as they can
have either sign. Because they contribute to only the quadratic and cubic
pieces of $V$, one can analyze their structure most readily along flat
directions in which the quartic pieces all vanish. Then, one finds two distinct
types of problems which may arise: potentials that are unbounded from below
(UFB), or potentials which break charge and/or color (CCB) at their minima.
Following Ref.~\cite{casas,vacuua1}, we will present a tree level based
classification of the various dangerous directions in various low energy
supersymmetric models with universal or non-universal soft terms.
\subsection{UFB cookbook}
In order to present the most dangerous of the UFB directions, it is useful to
notice some general properties:\footnote%
{In what follows $H_{1,2}^0 =H_{1,2},~H_1^- = h_1,~H_2^+ = h_2.$ In addition
$r,s,t$ stands for family indices and $i,j;a$ are $SU(2);SU(3)$ indices
respectively.}
Soft trilinear scalar terms $AY\tilde{F}H\tilde{F}^c$ cannot influence
dramatically a UFB direction, since for large enough values of the fields the
relevant $F$-terms dominate. The only negative contributions in $V^{(0)}$ that
can affect a UFB direction are:\footnote%
{Throughout, we use  the convention $m_3^2=\mu B < 0.$}
$m^2_2 |H_2|^2$, $-2|m_3^2| |H_1 H_2|$. Since these terms are quadratic, along
a UFB direction all the quartic positive terms coming from $F$- and $D$-terms
must be vanishing or kept under control.

\noindent\textbullet~~~\textbf{UFB-1} The first possibility is to play with
just $H_1,H_2$. This is the ordinary example of a UFB direction found in the
neutral Higgs potential along the direction $|H_1|=|H_2|$. As with all UFB
potentials, there is no quartic contribution to the potential along this
direction (a cubic piece is also absent here). In the present framework, since
we are mainly interested in the effect of one-loop corrections, we will also
allow the charged Higgs components to participate. This is calculationally
tractable only for this case contrary to the subsequent ones. Our choice for
the ``reduced'' function will be\footnote%
{With no harm of generality, we will assume real fields.}
\begin{eqnarray}
  V_{\mathit{UFB-1}}&=&V^{(0)}(H_1,h_1,H_2,h_2) +
  V^{(1)}(H_1,h_1,H_2,h_2),        \\[3mm]
   V^{(0)} &=& m_1^2 \lpa \magf{H_1} + \magf{h_1} \rpa +
   m_2^2 \lpa \magf{H_2} + \magf{h_2} \rpa + 2m_3^2
   \lpa H_1 H_2 - h_1h_2 \rpa         \nonumber \\
   &+& \frac{g_2^2}{2} \lbs H_1 h_2 + h_1 H_2 \rbs^2 +
   \frac{g^2+g_2^2}{8} \lpa \magf{H_1} + \magf{h_1} - \magf{H_2} -
   \magf{h_2} \rpa^2 \label{tree-UFB}
\end{eqnarray}
and $V^{(1)}$ necessary ingredients are presented in Appendix \ref{ufb1}.

\noindent \textbullet~~~\textbf{UFB-3} One other possibility is to take
$H_1=0$, $H_2 \neq 0$. We shall try to control the term $\lbs \mu H_2 \rbs^2 $
by introducing some squarks or sleptons. Notice that a nontrivial $\sUC_{ra}$
(specific $r,a$) will produce dominant $F$-terms, so it must be excluded. Using
the fact that $\sQU^T \equiv (~\squ~~\sqd~)$ and $\sLE^T
\equiv (~ \sne~~ \sle~)$, the remaining $F$-terms are%
\bedm
  \lbs Y_u^r \squ^a_r H_2 \rbs^2 + \lbs  Y_e^r \sne_r \sEC_r +
  Y_d^r \squ_r^a \sDC_{ra} \rbs^2 + \lbs \mu H_2 +
  Y_e^r \sle_r \sEC_r +  Y_d^r \sqd_r^a \sDC_{ra} \rbs^2
\endm
If we impose $\squ_r^a = 0$, then the non-trivial terms will become%
\bedm
  | Y_e^r \sne_r \sEC_r |^2 + \lbs \mu H_2 +   Y_e^r \sle_r
  \sEC_r +   Y_d^r \sqd_r^a \sDC_{ra} \rbs^2
\endm
Further suppression can be achieved in two ways:
\begin{description}
\item[$~~~\mathbf{a)}$]
    $\sEC_r=0,~~~\mbox{and}~~~\sDC_{ra}\neq 0~~~\mbox{and}~~~
      \sqd^a_r \neq 0~~~\mbox{for~a~specific~value~of~}r,a.$\\[2mm]
    In this case we do not eliminate quartic terms, but we try to neutralize
    their effect using additional scalars. It is not difficult to see that
    the strongest constraint arise for $r=3$ ($a$ whatever).
    An intuitive argument in favor of this is
    considering the soft squark contribution $m^2_{\sDC} |\sDC|^2$ to
    $V^{(0)}$. As we approach lower values of $V^{(0)}$, the relevant quartic
    $F$-term deactivation leads to $|\sDC|^2 \rightarrow |\mu H_2| / Y_d$. So,
    sbottom gives the less positive contribution. Control of $D$-terms also
    necessitates the presence of some slepton doublet provided that it does not
    introduce additional $F$-terms. $H_1$ or charged Higgs $h_1,h_2$ cannot play
    this role due to non-trivial quartic terms. Thus, the simplest choice which
    gives a minimal positive soft contribution and suppresses the quartic $D$-term
    $| H_2 \sle_r |^2$ is $\sLE^T_r \equiv (~ \sne_r~~0 ~ )\neq 0$ for some specific
    $r$. Our ``reduced'' function will then be
    \begin{eqnarray}
      V_{UFB-3a}&=& V^{(0)}(H_2,\sqd_3,\sDC_3,\sne_3)+
      V^{(1)}(H_2,\sqd_3,\sDC_3,\sne_3),~~~~(a=1)  \\[3mm]
      V^{(0)} &=& (m_2^2 - \mu^2) \magf{H_2} + m^2_{\sLE_3} \magf{\sne_3} +
      m^2_{\sQU_3} \magf{\sqd_3} + m^2_{\sDC_3} \magf{\sDC_3}   \nonumber \\%
      &+& \lbs \mu H_2 + Y_b \sqd_3 \sDC_3 \rbs^2 +\frac{g^2}{8} \lpa
      \magf{\sne_3} - \frac{1}{3} \magf{\sqd_3} -\frac{2}{3} \magf{\sDC_3} -
      \magf{H_2} \rpa^2         \nonumber  \\%
      &+& \frac{g_2^2}{8} \lpa \magf{\sne_3} - \magf{\sqd_3} - \magf{H_2} \rpa^2 +
      \frac{g_3^2}{6} \lpa \magf{\sqd_3} - \magf{\sDC_3} \rpa^2.
    \end{eqnarray}
\item[$~~~\mathbf{b)}$]
    $\sDC_{ra}=0,~~~\mbox{and}~~~\sle_r\neq 0~~~\mbox{and}~~~\sne_r=0~~~
      \mbox{and}~~~\sEC_r \neq 0~~~\mbox{for~some~specific~}r.$\\[2mm]
    Similarly as in the case (a), the strongest constraint comes for $r=3$.
    Further control of $D$-terms requires a slepton doublet from another
    family ($h_2$ is inappropriate since it introduces quartic terms). So we take
    $\sLE^T_s \equiv (~ \sne_s~~0 ~ )\neq 0$ and $\sEC_s=0$ for some specific
    $s\neq r$. Choosing $s=2$, we get the following ``reduced'' function%
    \begin{eqnarray}
    V_{\mathit{UFB-3b}}&=& V^{(0)}(H_2,\sle_3,\sEC_3,\sne_2)
      + V^{(1)}(H_2,\sle_3,\sEC_3,\sne_2),    \\[3mm]
      V^{(0)} &=& (m_2^2 - \mu^2) \magf{H_2} + m^2_{\sLE_3} \magf{\sle_3} +
      m^2_{\sLE_2} \magf{\sne_2} + m^2_{\sEC_3} \magf{\sEC_3}   \nonumber \\%
      &+& \lbs \mu H_2 + Y_{\tau} \sle_3 \sEC_3 \rbs^2 +\frac{g^2}{8} \lpa
      \magf{\sne_2} + \magf{\sle_3} -2 \magf{\sEC_3} - \magf{H_2} \rpa^2
      \nonumber  \\
      &+& \frac{g_2^2}{8} \lpa \magf{\sne_2} - \magf{\sle_3} - \magf{H_2} \rpa^2
    \end{eqnarray}
    while $V^{(1)}$ pieces are shown in Appendix \ref{ufb3b}.
\end{description}

\subsection{CCB cookbook}
CCB most readily occurs along directions which are $D$-flat, though not
necessarily $F$-flat (the $\phi^4$ contributions to the potential are
suppressed by Yukawas). In order to gain intuition about CCB, let us present
some general comments: The most dangerous CCB directions involve \emph{only
one} particular trilinear coupling of one generation. Two or more trilinear
couplings with different values of $Y$ can not interfere constructively in the
same region of field space to deepen the potential. The CCB directions we
explore are not $F$-flat, so the most restrictive are those with the smallest
non-vanishing $F$-terms. Since they are proportional to the respective Yukawa
coupling, we expect cases with the lightest $Y$ to be more restrictive. The
most dangerous CCB directions are:

\noindent \textbullet~~~\textbf{CCB-E\textit{a}} Our goal here is to keep only
the ``light'' $A_e$ trilinear coupling. For that reason we assume $ H_1 \neq
0$, $\sEC_r \neq 0$, $\sLE_r^T = ( ~ \sne_r~ \sle_r ~ ) \neq 0$ (for some
specific $r$). The relevant $F$-terms are%
\bedm
  | Y_e^r \sne_r \sEC_r |^2 + | Y_e^r \sle_r \sEC_r |^2
  + | \mu H_1 |^2 + | Y_e^r H_1 \sEC_r |^2
  + | Y_e^r \sle_r H_1 |^2
\endm
Obviously further suppression is possible so we choose $\sne_r = 0$ (the case
$\sle_r$ = 0 kills the $A_e$ trilinear term) and introduce some squarks
(sleptons will add a new $A$) to control the pure Higgs $F$-term. We can take
$\sQU_s^{aT} = ( ~ \squ_s^a ~ \sqd_s^a ~ ) \neq 0$, and $\sUC_{sa} \neq 0$ (for
some specific $s,a$). Then the relevant $F$-term becomes \bedm
  \lbs \mu H_1 - Y_u^s \squ_s^a \sUC_{sa} \rbs^2 + | Y_u^s \sqd_s^a
  \sUC_{sa} |^2
\endm
and the minimal choice is for $\sqd_s^a = 0$. Using a similar intuitive
argument as in \textbf{UFB} case, we conclude that $Y_t$ gives the smaller
positive $F$-term. Choosing $r=1$ (lighter Yukawa), $s=3$, $a=1$ makes further
control of $D$-terms unnecessary. Putting altogether, the ``reduced'' function
becomes
\begin{eqnarray}
  V_{\mathit{CCB-Ea}} &=& V^{(0)}(H_1,\sle_1,\sEC_1,\squ_3,\sUC_3) +
  V^{(1)}(H_1,\sle_1,\sEC_1,\squ_3,\sUC_3),       \\[3mm]
  V^{(0)} &=& (m_1^2 - \mu^2) \magf{H_1} + m_{\sEC_1}^2 \magf{\sEC_1}
  + m_{\sLE_1}^2 \magf{\sle_1} + m_{\sQU_3}^2 \magf{\squ_3}
  + m_{\sUC_3}^2 \magf{\sUC_3}   \nonumber \\
  &+& 2 Y_e A_e \sle_1 \sEC_1 H_1 + \lbs \mu H_1 - Y_t \squ_3 \sUC_3
  \rbs^2 + | Y_e \sle_1 \sEC_1 |^2 + | Y_e \sEC_1 H_1 |^2 +
  | Y_e \sle_1 H_1 |^2   \nonumber   \\
  &+& \frac{g^2}{8} \lpa \magf{\sle_1} - 2 \magf{\sEC_1}  - \frac{1}{3}
  \magf{\squ_3} + \frac{4}{3}\magf{\sUC_3} + \magf{H_1} \rpa^2
  \nonumber  \\
  &+& \frac{g_2^2}{8} \lpa \magf{\sle_1} - \magf{\squ_3} - \magf{H_1}
  \rpa^2 + \frac{g_3^2}{6} \lpa \magf{\squ_3} - \magf{\sUC_3} \rpa^2.
\end{eqnarray}

\noindent \textbullet~~~\textbf{CCB-Eb} The other way to keep $A_e$ is by two
Higgses $H_1 \neq 0, H_2 \neq 0$. Now one should allow the same sleptons
$\sEC_r \neq 0$, $\sLE_r^T = ( ~ \sne_r ~ \sle_r  ~ ) \neq 0$ (for some
specific $r$). However, here we cannot use squarks or sleptons to compensate
the $H_1$ $F$-term, because this will introduce an additional trilinear
coupling. The relevant $F$-terms will be \bedm
  | Y_e^r \sne_r \sEC_r |^2 + \lbs Y_e^r \sle_r \sEC_r+ \mu H_2 \rbs^2
  + | \mu H_1 |^2 + | Y_e^r \sEC_r H_1 |^2 + | Y_e^r \sle_r H_1 |^2.
\endm
By taking $\sne_r = 0$, we ensure minimal $F$-terms without killing the
trilinear contribution. In the lighter case ($r=1$), further control of
$D$-terms is not necessary, so our ``reduced'' function will be
\begin{eqnarray}
  V_{\mathit{CCB-Eb}} &=& V^{(0)}(H_1,H_2,\sle_1,\sEC_1) +
  V^{(1)}(H_1,H_2,\sle_1,\sEC_1),       \\[3mm]
  V^{(0)} &=& m_1^2 \magf{H_1} + m_2^2 \magf{H_2} + m_{\sLE_1}^2 \magf{\sle_1}
  + m_{\sEC_1}^2 \magf{\sEC_1} + 2 m_3^2 H_1 H_2   \nonumber  \\
  &+&  2 Y_e \sle_1 \sEC_1 (\mu H_2 + A_e H_1)
  + | Y_e \sle_1 \sEC_1 |^2 + | Y_e \sEC_1 H_1 |^2
  + | Y_e \sle_1 H_1 |^2      \nonumber  \\
  &+& \frac{g^2}{8} \lpa \magf{H_1} + \magf{\sle_1} -2 \magf{\sEC_1}
  - \magf{H_2} \rpa^2 + \frac{g_2^2}{8} \lpa \magf{H_2} + \magf{\sle_1}
  - \magf{H_1} \rpa^2 \label{colbr}
\end{eqnarray}
Expressions of $V^{(1)}$ are presented in Appendix \ref{ccbeb}.
\section{\label{calsch}Calculational Scheme}

Let us now describe the method used to probe the dangerous directions just
presented. As is well known, one-loop effective potential depends on the
eigenvalues of the tree level mass matrices. To be more accurate, these
corrections are derived under the assumption that the potential is resting at a
tree level minimum (eigenvalues of $\partial_{ij}V^{(0)}$ positive). However,
in a UFB case we need to define the one-loop corrections away from a classical
minimum. Hence several eigenvalues of the scalar mass matrix may be negative
leading to a complex valued function. These contributions are a signal that the
sum of one-particle irreducible diagrams $V_{\rm 1PI}$ does not give the
effective potential. Formally, in a case like that we must use the convex
envelope of $V_{\rm 1PI}$, which in the vicinity of a classical minimum matches
the usual loop expansion. Trying to estimate the one loop corrections despite
non-convexity of the effective potential, we adopt a moderate approach, namely
$V^{(1)}$ everywhere is given by the real part of the ordinary one-loop
expressions \cite{complex,grz}. Thus, in a mass-independent renormalization
scheme ($\overline{\rm DR}$) \cite{siegel} $V^{(1)}$ is given by ($\tilde{Q}=Q
e^{3/4}$)
 \bea  \label{v1}
    V^{(1)} &=& k \sum_{\stackrel{\scriptstyle p}{(M_p^2 \not=0)}}V_1^{(p)}
    ~~~\mbox{with}~~~
    V_1^{(p)}= \frac{(-1)^{2 S_p}}{4} (2 S_p + 1) \, C_p \, \mathcal{N}_p
          \, M_p^4(\phi) \ln{ \frac{|M_p^2(\phi)|}{\tilde{Q}^2}}
 \ena
where $k=(16 \pi^2)^{-1}$. $M_p$ stands for the tree level mass eigenvalue of
the ${\rm p}^{\rm th}$ particle, $\mathcal{N}_p$ is the number of its helicity
states and the associated spin, color degrees of freedom are denoted by $S_p$,
$C_p$ respectively. It is evident from (\ref{v1}) that in the case of a single
mass scale a judicious choice of the renormalization scale ($Q^2 = M^2(\phi)$)
eliminates all large logs. However, in the case of many different mass scales
any renormalization scale will leave large logs remnants behind, so that we
need higher loop corrections in the loop expansion to trust the results.

The heart of the problem lies in the renormalization scheme we use
($\overline{\rm DR}$). For a mass independent scheme the decoupling of various
mass states is not automatic and has to be incorporated. Hiding all heavy
particle loop contributions in a redefinition of some lower energy parameters,
we secure that all masses smaller than a given scale (decoupling scale) behave
as massless, while larger masses decouple and never generate problems in
perturbation theory. Below a decoupling scale the theory is an effective field
theory with new RGEs, while threshold effects take care of the matching between
both theories at the boundary scale.

A simple way to realize this scenario is to treat the thresholds as steps in
the particle content of the RGE $\beta$-functions \cite{sak}. Usually one
integrates the RGEs from a superlarge scale $M_X$ to any desirable value of
$Q$. As we come down from $M_X$, as long as we are at scales larger than the
heaviest particle threshold, we include contributions from all particles in the
model. When we cross the heaviest particle threshold, we switch to a new
effective field theory with the heaviest particle integrated out and of course
a new $\beta$. For field configurations in the low energy regime ($\lesssim
300$ GeV), the condition to determine the exact point of decoupling is simply
$\tilde{Q}^2 = |m^2(Q)|$, where $m^2(Q)$ is the running soft parameter
corresponding to the particle.\footnote
{We use the factor $e^{3/4}$ for compatibility with Eq.~(\ref{v1}).}
Obviously, the step functions in RGEs will have the form $\theta_m = \theta
(\tilde{Q}^2 - |m^2(Q)|)$. Alternatively, for all other field configurations,
the decoupling points are fixed by $\tilde{Q}^2=|M^2(\phi;Q)|$, where $M^2$ is
the field dependent mass eigenvalue of a particle. Analogously, the step
functions in RGEs will now become $\theta_M = \theta(\tilde{Q}^2 - |M^2(\phi;Q)
|)$, where $\phi$ are in principle all the relevant fields along the dangerous
direction under consideration. Additionally, knowing that the minimum of
$V^{(0)}$ lies at a non trivial field configuration, a suitable generalization
of vacuum subtraction in
Eq.~(\ref{1st}) is given (at one-loop order) by \cite{omega}%
 \be
 -\Omega^{'} = V^{(0)}(\lg H_1 \rg, \lg H_2 \rg; \blm{0}) +
               V^{(1)}(\lg H_1 \rg, \lg H_2 \rg, \blm{0};Q)
\label{Omega}
\ee%
Finally, we also replace the potential (\ref{v1}) by \cite{quir}%
 \be
     V^{(1)} = k \sum_i V_i^{(1)} \theta_i  ~~~~~~\mbox{where}~~~~~~
   \theta_i \equiv \theta\Big(\tilde{Q}^2 - |M_i^2(\phi;Q)|\Big)
   \label{vth}
\ee%
One other issue is that of choosing the renormalization point. Following
\cite{dvg}, we use a prescription that  preserves the hierarchy ($|V^{(0)}|
\gtrsim |V^{(1)}|)$ and has a continuous field dependence
\be%
\tilde{Q}^* = 10^{\omega(x)} \sqrt{\mathbf{\Phi}  \cdot
\mathbf{\Phi}},~~~~~~~~~~~~~~~~\mathbf{\Phi}\cdot\mathbf{\Phi}= \sum_i \phi_i^2
\label{Qstar}
\ee%
where $x=\log (\sqrt{\mathbf{\Phi} \cdot \mathbf{\Phi}}/\Lambda)$ ($\Lambda =
1$ GeV makes the log argument dimensionless), $\phi_i$ are the scalar fields
involved in the dangerous direction under consideration and our ansatz for
$\omega(x)$ is given in Appendix \ref{omega}.

We stress here that not only the effective potential in Eq.~(\ref{vth}), but
also the $\beta$ functions in RGEs do depend on the same step functions
 \be
 \frac{d\xi}{d\ln Q} =
    \beta_\xi \left(\theta_i(\tilde{Q}^2-|M_i^2(\phi;Q)|)\right)
 \label{ksi}
 \ee
Formally, we can integrate these implicitly field dependent RGEs to the
appropriate renormalization scale $\tilde{Q}^*$ and then construct the required
effective potential. However, from a practical point of view using all the
relevant scalar fields involved in Eq.~(\ref{ksi}) requires excessive
computational time. The computation of the field dependent mass eigenvalues
$M_i^2(\phi;Q)$ is very complicated due to higher dimensional mass matrices and
has to be repeated for each integration step taken internally by our numerical
integrator. A possible way out is to restrict ourselves to the neutral Higgs
subset of $\phi$'s \emph{solely} in Eq.~(\ref{ksi}). In that case the mass
eigenvalues are trivially given by solving algebraic equations, but the most
important is that the computational time is reduced by several times. We stress
here that this kind of approximation is applied only to the $\beta$ functions
in RGEs (Eq.~(\ref{ksi})) and not to the effective potential (Eq.(\ref{vth}))
otherwise loop expansion is in danger. We have also checked the results of the
two alternatives just described for some representative points of the parameter
space and found no significant difference. So, finally in our investigation we
have adopted the second way which is the quickest.

To complete the picture, one also needs some ``boundary scale''
$Q_{\mathit{high}}$ where the starting values of the running parameters
(couplings, fields) should be provided for the evolution at $Q^*$. Notice that
now, due to field dependent thresholds in $\beta$ functions, the RGE evolution
for $Q < Q_{\mathit{high}}$ depends on the field point we are. A convenient
choice for $Q_{\mathit{high}}$, besides the unification scale, is an
intermediate scale higher than the largest field dependent mass eigenvalue at
the current field point. Valid choices for $Q_{\mathit{high}}$ are
$Q_{\mathit{high}} \gtrsim 2.12 \phi_\infty$, where $\phi_\infty$ stands for
the upper bound order of magnitude of the allowed values for the scalar fields
\cite{dvg}. Besides, we also need to know the values of the running parameters
and fields there. For the fields the most plausible option is to take
$\phi(Q_{\mathit{high}}) = \phi_0$ where $\phi_0$, is the field point we
examine. Since $Q_{\mathit{high}}$ is above all thresholds, the required values
for the couplings there should not depend on the background fields and a
reasonable choice is%
\be
 \lambda_\alpha(Q_{\mathit{high}};\blm{\phi_0})=\lambda_\alpha(Q_{\mathit{high}};\lg
 H_1 \rg, \lg H_2 \rg, \blm{0})
\ee%
where the RHS is obtained by running \cite{sak} the couplings from their known
values at $M_Z$ when the potential is resting at its physical vacuum. Evolving
this set of values $\{\blm{\phi_0},~\lambda_\alpha (Q_{\mathit{high}};\lg H_1^0
\rg, \lg H_2^0 \rg,\blm{0})\}$ from $Q_{\mathit{high}}$ to $Q^*$ using field
dependent thresholds, the effective potential at the current field point can be
constructed.

Minimization of the effective potential is performed numerically. Since
analytic expressions for the derivatives are not available, we resort to
methods that require only function evaluations. One such very efficient method
is the downhill Simplex method \cite{simplex}. A simplex (or polytope) in $n$
dimensional Euclidean space is a construct with $n+1$ vertices defining a
volume element. For instance, in two dimensions the simplex is a triangle, in
three dimensions is a tetrahedron and so on. Using a population of $n+1$ points
(simplex vertices), the algorithm brings the simplex in the area of a minimum
and adapts it to the local geometry. The initial Simplex may be constructed
from the current point (first vertex) by taking a single step along each of the
$n$ dimensions.
\section{\label{resl}Results}

Using the procedure outlined above and Merlin package \cite{merlin}, we
explored regions of MSSM parameter space for unphysical vacuua. We begin our
discussion for the allowed parameter space in the ($m_0, m_{1/2}$) plane.
\begin{figure}[t!]
\centering
   \begin{minipage}[c]{0.5\textwidth}
   \vspace*{-10mm}\centering \includegraphics[width=0.9\textwidth]{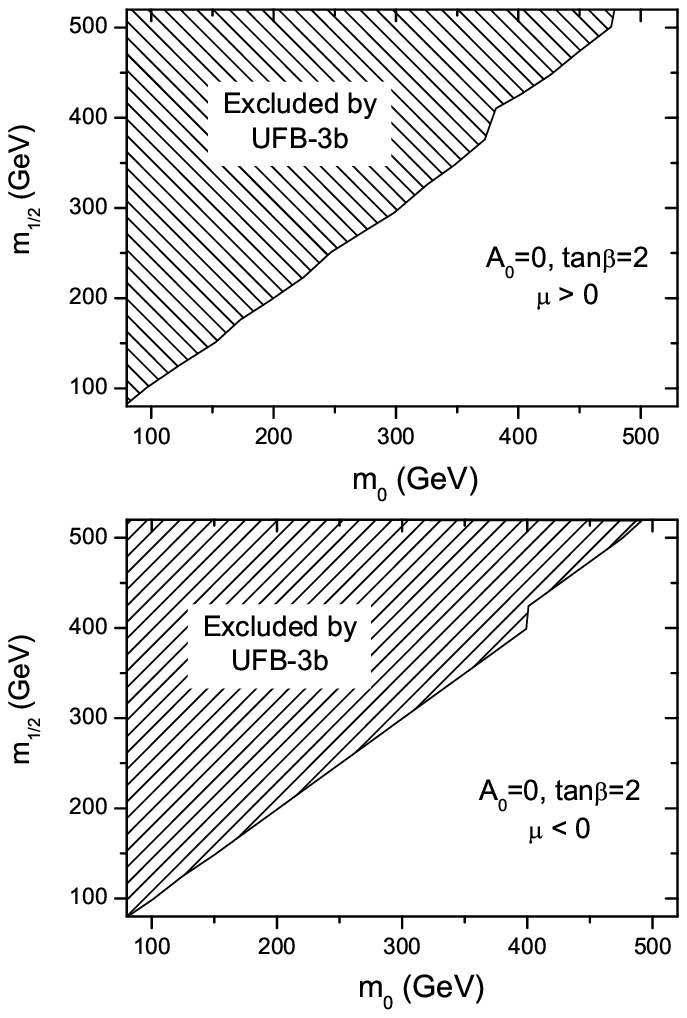}
   \end{minipage}%
   \begin{minipage}[c]{0.5\textwidth}
   \vspace*{-10mm}\centering \includegraphics[width=0.9\textwidth]{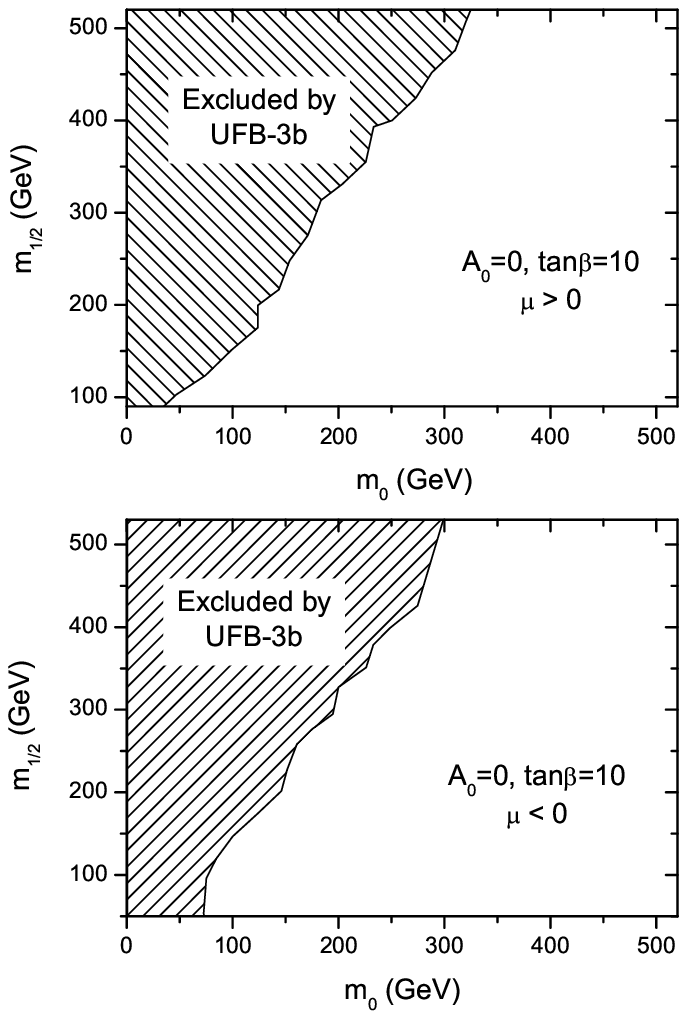}
   \end{minipage}%
\caption{Exclusion plots in the $m_0$ vs $m_{1/2}$ plane for the UFB-3b
direction} \label{gr1}
\end{figure}%
\begin{figure}[b!]
\centering
\begin{minipage}[c]{0.5\textwidth}
\centering \includegraphics[width=0.9\textwidth]{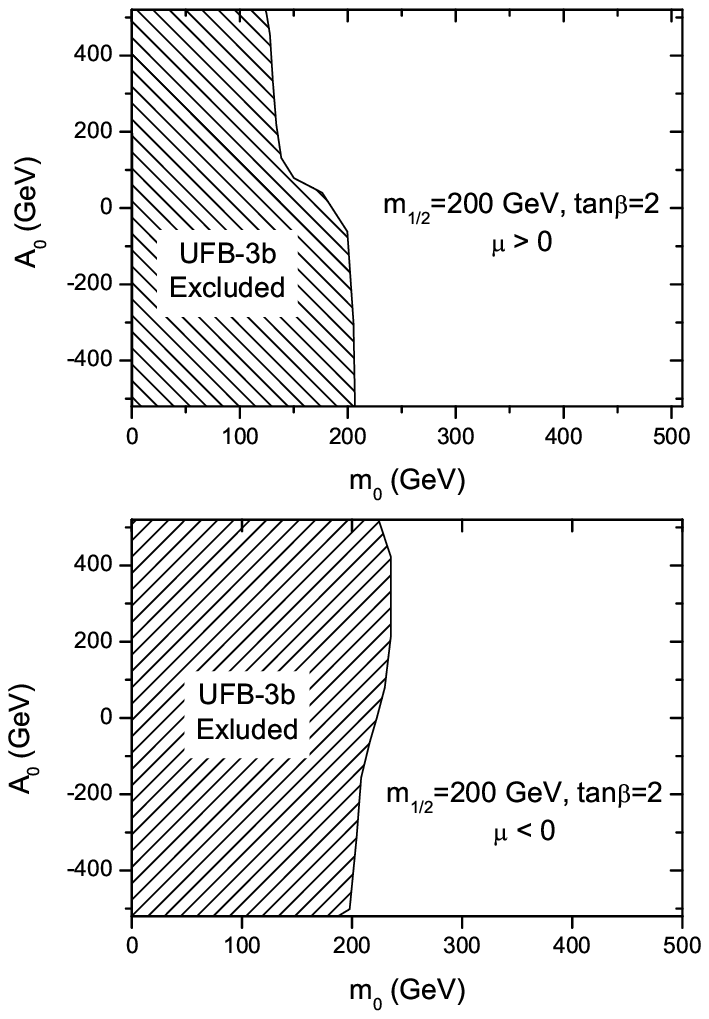}
\end{minipage}%
\begin{minipage}[c]{0.5\textwidth}
\centering \includegraphics[width=0.9\textwidth]{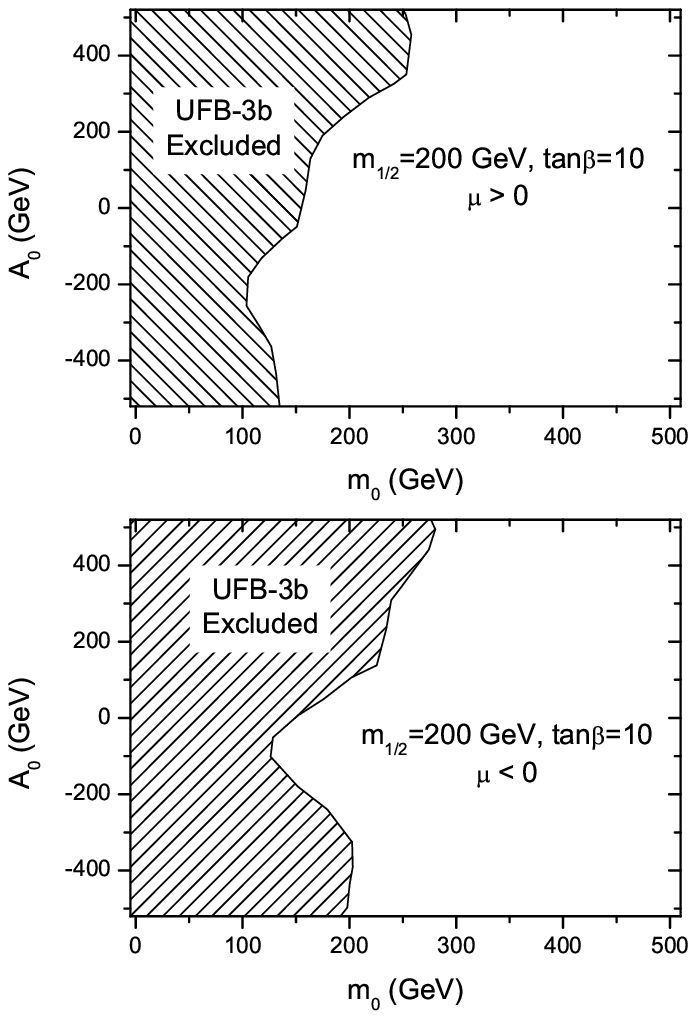}
\end{minipage}%
\caption{Exclusion plots in the $m_0$ vs $A_0$ plane for the UFB-3b
direction}\label{gr2}
\end{figure}%
We fix $\tan\beta$ to be $2$ or $10$, and take $A_0=0$. Our search was
performed in the range $(0,500)$ GeV for both axis and the grid was scanned
with $25$ GeV resolution. In Fig.~\ref{gr1} the shaded regions to the left of
the solid line display points where UFB-3b escapes were discovered. These
parameter values with $m_0$ smaller than $m_{1/2}$ indicate that much of the
mSUGRA parameter space associated with light sleptons is ruled out. It is also
worth noting that higher values of $\tan\beta$ weaken these constraints.

Although Fig.~\ref{gr1} is plotted for $A_0=0$, a similar excluded region
results for other choices of the $A_0$ parameter. An analogous plot in
Fig.~\ref{gr2} displays the effect of variation in the trilinear $A_0$
parameter in combination with $m_0$ for fixed $m_{1/2}$ at $200$ GeV. Again, as
usual, we fixed $\tan\beta$ to be $2$ or $10$. Performing a numerical
minimization along the UFB-3b direction, unphysical configurations were found
only for initial values taken from the shaded regions. Similarly, we find out
that the vacuum constraints disfavor light slepton cases.

\begin{figure}[t!]
 \centering
\begin{minipage}[c]{0.5\textwidth}
\centering \includegraphics[width=0.9\textwidth]{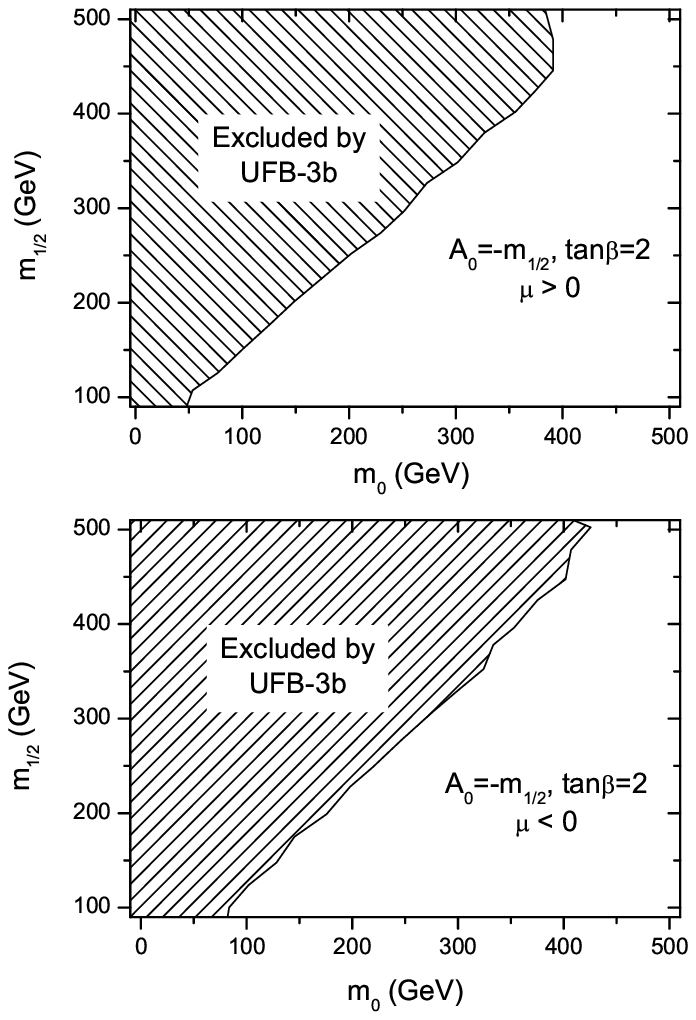}
\end{minipage}%
\begin{minipage}[c]{0.5\textwidth}
\centering \includegraphics[width=0.9\textwidth]{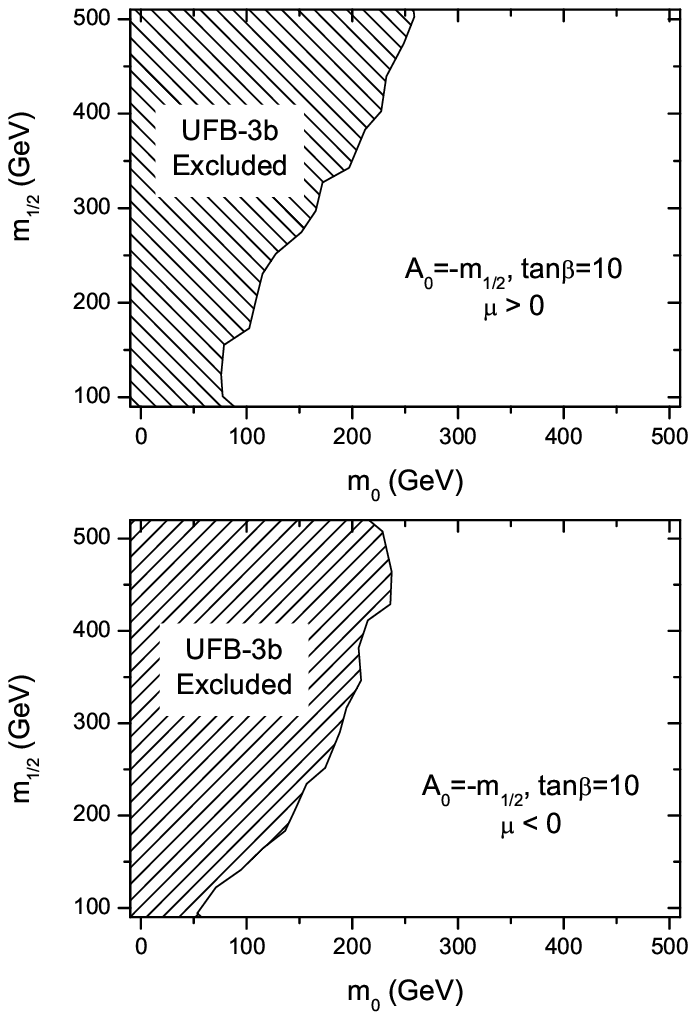}
\end{minipage}%
\caption{Exclusion plots in the $m_0$ vs $m_{1/2}$ plane with the additional
constraint $A_0=-m_{1/2}$ for the UFB-3b direction}\label{gr3}
\end{figure}

We have also examined a more general non-universal case coming from string
models, which encompasses the special case where supersymmetry is broken in the
dilaton sector. In the latter case one is led to GUT scale soft terms related
by $m_{1/2} = - A_0 = \sqrt{3} m_0$. However, in our examination we treat $m_0$
as a free parameter. We performed scans in the ($m_0, m_{1/2}$) plane with the
usual values for $\tan\beta$ ($2$ or $10$). Numerical minimization showed that
for small $m_0$ much of the parameter space is excluded by the UFB-3b
constraint, as shown in Fig.~\ref{gr3}. Furthermore, it is easy to see that the
so-called dilaton dominated scenario corresponds to a straight line located
entirely inside the forbidden region.

A similar scan in the parameter space of the MSSM has also been performed along
the dangerous direction CCB-Eb. We considered again the three different regions
depicted in Figs \ref{gr1}-\ref{gr3}. Using the simplex minimization procedure
and the one-loop expressions in Appendix \ref{ccbeb}, we found no dangerous CCB
minima. Tree level constraints described in Appendix \ref{th_ccb} have also
considered, as the numerical procedure scanned various field configurations and
found that they are not violated for fields $|\phi| \sim A_e/Y_e \sim
10^3/10^{-5}$ GeV $\sim 10^8$ GeV. In great measure this is due to the large
and positive contribution to the potential of the soft masses and especially
$m_1^2$. Let us stress here that our renormalization scale choice depends on
the field point we are. If at some field point $\phi_0$ the renormalization
scale $Q^*_0=Q^*(\phi_0)$ does reverse the inequality of a tree level
constraint, then the tree level potential constructed from ``soft'' parameters
at the specific scale $Q^*_0$ will be deeper than the physical vacuum for
fields $|\phi_d| \sim A_e(Q^*_0) / Y_e(Q^*_0) $. In other words, tree level
constraints signal an unphysical situation only if they are violated for fields
of the order $ \sim
A_e(Q^*)/Y_e(Q^*)$.%
\begin{figure}[ht]
 \centering
\begin{minipage}[c]{0.5\textwidth}
\centering \includegraphics[width=0.9\textwidth]{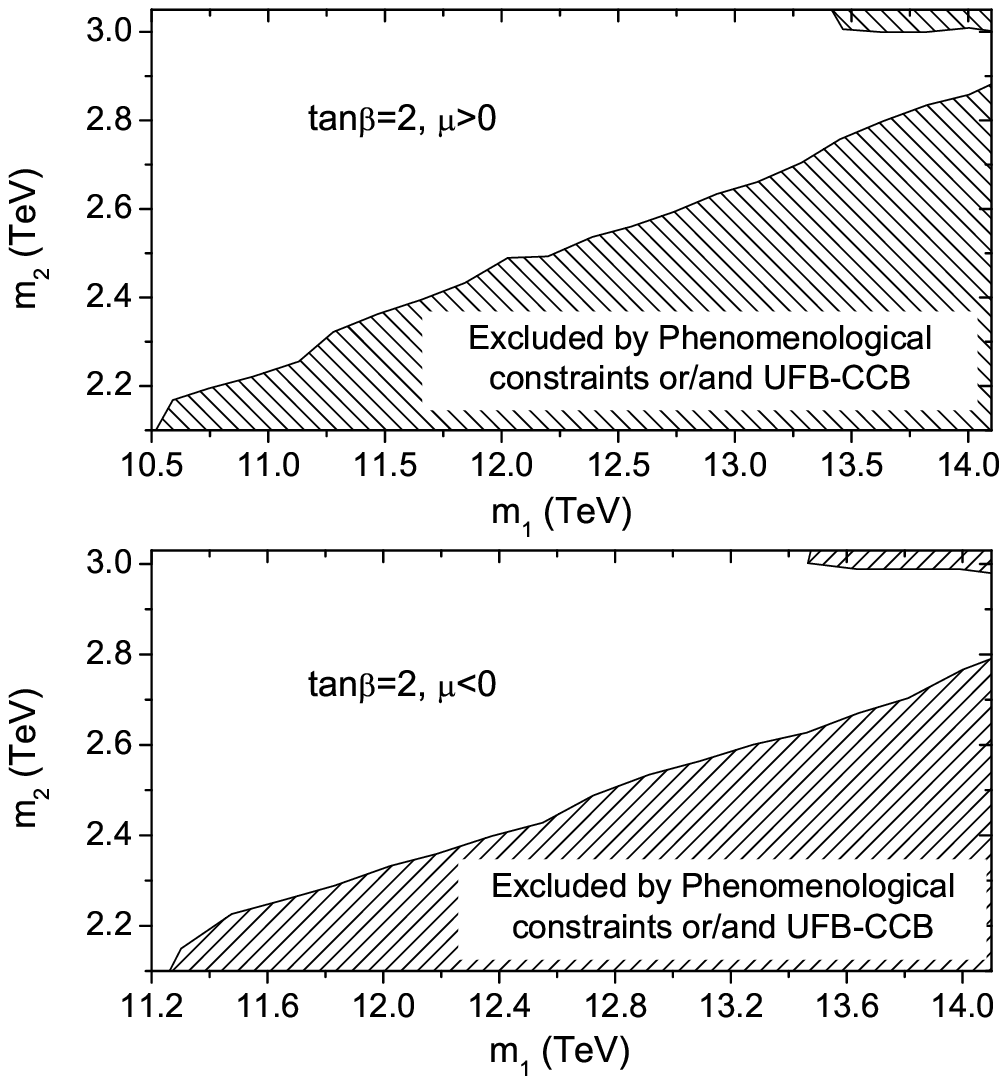}
\end{minipage}%
\begin{minipage}[c]{0.5\textwidth}
\centering \includegraphics[width=0.9\textwidth]{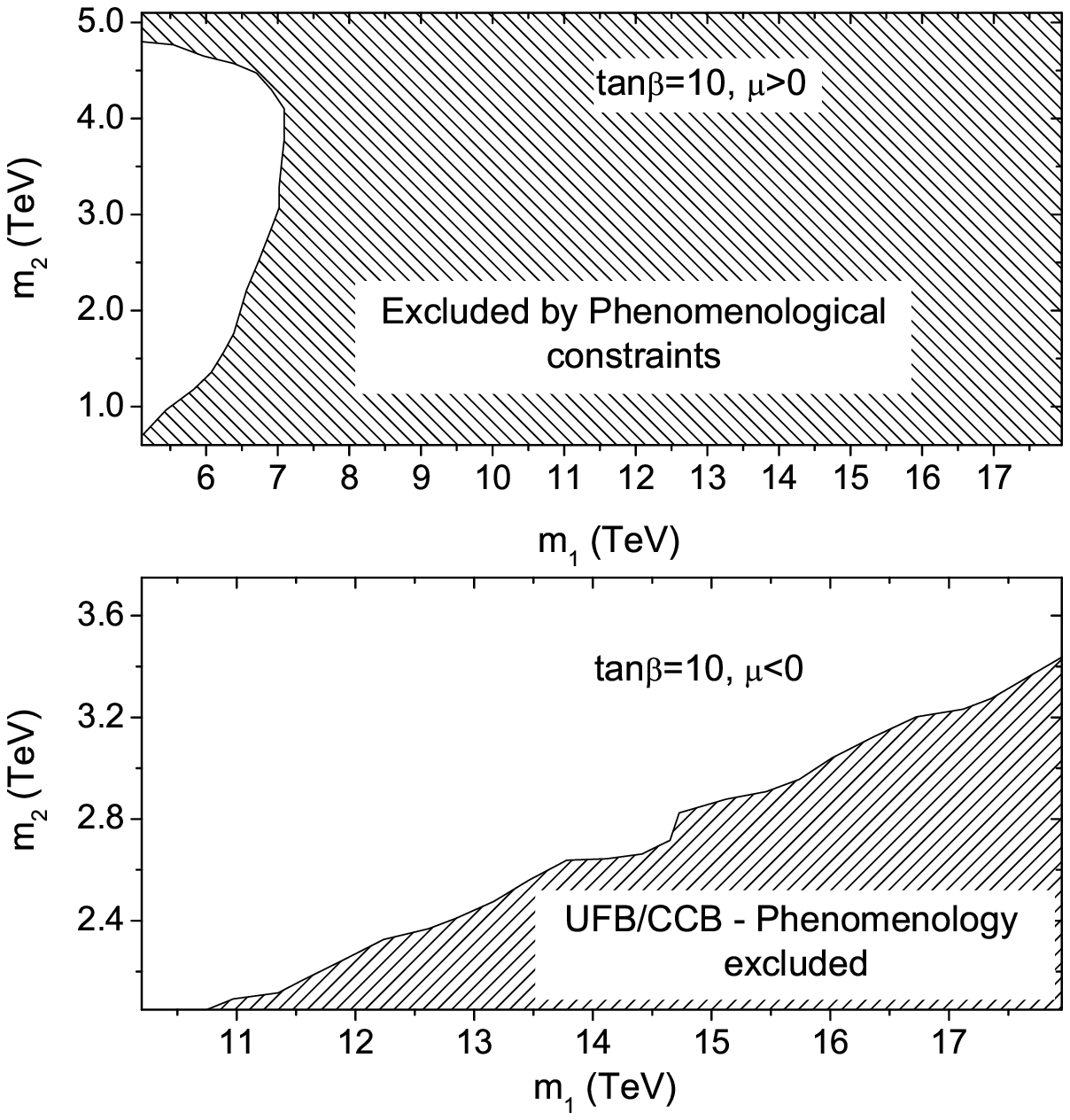}
\end{minipage}%
\caption{Exclusion plots for the $m_1$ vs $m_2$ plane in a brane world
model.}\label{gr4}
\end{figure}

We next focus our attention on the non-universal brane model and perform a scan
for unphysical vacuua.  In that case, due to non-universality, sparticles may
generate too large flavor-changing (FC) effects but the specific brane model
offers a mechanism for suppressing these dangerous processes. In this scenario,
pushing the SUSY breaking scale at the first brane high enough ($m_1 \simeq 10
- 20$ TeV), the masses of the first and second generations scalars will become
sufficiently heavy to avoid the flavor problem. On the other hand, introducing
a second SUSY breaking scale ($m_2$) attached to the second brane constrains
the heavy scalars to be not much heavier than $1$ TeV or so to preserve the
gauge hierarchy. We have performed scans in the $(m_1, m_2)$ plane for the
values described above. Our results for $\tan\beta = 2, 10$ are shown in
Fig.~\ref{gr4}, where one can also see the regions excluded by phenomenological
constraints\footnote%
{By this we mean either no electroweak breaking or no mass for the gluino or
violation of the experimental bounds for superpartners masses}
or/and by unphysical vacuua driven radiatively by negative scalar mass
eigenvalues. Note that in the case $\tan\beta = 10,~\mu>0$ the parameter space
is strongly restricted allowing only a very small ``window'' for phenomenology.

Unfortunately, an analytical treatment of these issues is tedious and leads to
very unwieldy expressions \cite{ferr}. Relaxing our preciseness requirements,
we can intuitively realize the strength of a dangerous direction by resorting
to the familiar tree level expressions. Constraints based on tree level logic
have been thoroughly studied in the past \cite{casas,vacuua1,vacuua3} and our
intention here is not to reproduce these analyses, but merely to use them as an
alternative confirmation since our renormalized effective potential respects
perturbation series hierarchy. Indeed, as one can see in Appendix \ref{th_ufb},
the global minimum of Eq.~(\ref{tree-UFB}) (UFB-1 case) is not deeper than the
physical vacuum, a fact that has been also numerically justified at one-loop
level.

One way to test reliability of the effective potential is by checking scale
invariance \cite{grz,ferr,mart}. In our case the chosen scale\footnote{
Note that our renormalization point is field dependent rendering $\Omega'$ in
Eq.~(\ref{Omega}) a non trivial subtraction}
can not vary arbitrarily, since this endangers convergence of loop expansion.
For very large values of scale $\theta$ steps in $V^{(1)}$ allow all heavy
masses and loop corrections become huge. We have tried several scale
prescriptions (various $\omega$ choices) and found that only the location of a
UFB-3b escape is sensitive to scale variations. We believe that this
sensitivity is due to cancellations that take place along a UFB-3b escape and
higher loop effects. Since our expressions are only one-loop scale invariant,
the higher loop difference is amplified by these cancellations and appears as a
small ``rotation'' of the escape trajectory. The important point is that the
physical picture remains unchanged: whatever $\omega(x)$ one chooses (within
perturbativity constraints) cannot render an unstable potential stable and vice
versa.

In the present work we mainly focus on an explicit treatment of the effect one
loop radiative corrections may have on various UFB and CCB dangerous directions
in supergravity models. Due to the intrinsic complexity of this subject, as
well as the theoretical problem of multiple mass scales these issues have not
been straightly touched upon in the past. Here for the first time, implementing
a sophisticated threshold technique \cite{quir,dvg} outlined in
Sec.~\ref{calsch}, we were able both to overcome the multiscale problem and
deal explicitly with the radiative corrections. We believe that this is a more
complete approach leading to more reliable results and corresponding plots.
Practically, our improved results are in broad agreement with previous
investigations \cite{vacuua1,vacuua3}. However, a more detailed comparison is
highly non-trivial due to the variety of methods implemented by the various
authors.
\section{\label{conc}Conclusions}
Minimal SUGRA scenarios provide a well motivated and phenomenologically viable
framework of how weak scale supersymmetry might occur. Unfortunately, one has
to pay the price of introducing extra degrees of freedom and new parameters
leading to the appearance of new sources of flavor changing and CP violating
processes, large number of new parameters and complicated scalar sector of the
theory. On the other hand, the way low energy is related to the fundamental
theory may shed some light to the solution of these problems.

In this paper, trying to constrain the parameter space of some mSUGRA models as
described in introduction, we follow a strategy of finding regions in the
general scalar field configuration space which have inferior values than the
physical vacuum. The argument is that, if this configuration is not a minimum,
an unbounded from below direction or a global minimum with charge and/or color
broken surely exists somewhere and the associated point in parameter space
should be excluded from consideration.

We have analyzed the relevant potentials employing the full one-loop radiative
corrections in the calculation. Because of the various mass scales present in
these models, renormalization group improvement of the potential has
ambiguities and should be carefully treated. Implementing the decoupling
theorem in a manner proposed by the authors of Ref.~\cite{quir,dvg}, we treat
the various particle thresholds as steps in the $\beta$-functions as well as in
the one loop corrections of the scalar potential. We stress here the role
played by the renormalization scale choice, as given in Sec.~\ref{calsch}\@. It
should be wisely chosen in order to eliminate heavy particles whose
participation puts in danger the convergence of the loop expansion
(\ie~$|V^{(1)}| \lesssim |V^{(0)}| $).

Employing the framework just stated and using numerical minimization procedure,
we have performed an analysis of how the most dangerous directions put
restrictions on the whole parameter space of the various models described. In
both the MSSM case and the brane world scenario considered, these constraints
turn out to be very strong, producing important bounds not only on the value of
$A$ (soft trilinear coupling), but also on the values of $m_{1/2}$ (gaugino
masses) and $m_0$ (scalar masses). Our analysis is summarized in Figs 1-4. As a
general trend, the smaller the value of $m_0$ the more restrictive the
constraints become.

Finally, we note that the more or less strict constraints derived here from
non-standard vacuua are avoided if we adopt the idea \cite{lang} that we may
indeed exist in a false vacuum and the tunneling rate from our present vacuum
to a non-standard one might be small relative to the age of the universe.
However, it is difficult to realize the circumstances under which the idea that
we may live in a false, metastable vacuum could be reconciled with the
existence of a small positive dark energy in the universe, either in the form
of a constant vacuum energy/cosmological constant or in the form of a new
scalar quintessence field, as the recent discoveries suggest \cite{perl}.
\begin{acknowledgments}
We would like to thank J.~Rizos for discussions. DG would also like to thank
D.~Katsanos for his help with numerical routines. We also wish to thank the
anonymous referee for his constructive comments which have led to an improved,
more clarified presentation. This research is partially supported by European
Union under the contract RTN No HPRN-CT-2000-00152.
\end{acknowledgments}
{                     
\appendix%
{\centering \section{\label{th_ufb}Analytic treatment of UFB-1 at tree level}}

In this appendix we compute the global minimum of the tree level effective
potential along the UFB-1 direction. Assuming for simplicity real fields, the
solution is based on the new ``polar'' variables $(R_1, R_2, \varphi, \theta)$
where \be
   H_1  = R_1 \cos \varphi \qquad
   h_1  = R_1 \sin \varphi \qquad
   H_2  = R_2 \cos \theta \qquad
   h_2  = R_2 \sin \theta
\ee%
Of course $R_1 >0, R_2>0$ for an invertible transformation, thus the trivial
case should be examined separately. If we take both or at least one $R$
trivial, we also find a trivial value for the potential extremum which is
clearly above the physical vacuum shown in Eq.~(\ref{physv}). Hence it is
sufficient to
minimize the function%
\be%
V^{(0)}  = m_1^2 R_1^2  + m_2^2 R_2^2  + 2m_3^2 R_1 R_2 \cos (\varphi  + \theta
) + \frac{{g_2^2 }}{2}R_1^2 R_2^2 \sin ^2 (\varphi  + \theta ) + \frac{{g^2  +
g_2^2 }}{8}(R_1^2  - R_2^2 )^2
\ee%
Solution of the minimization conditions leads to the following cases:
\begin{description}
  \item[(I)]$R_1 R_2 \cos(\varphi+\theta)=2m_3^2/g_2^2,~~~\partial_{R_1}V^{(0)}=0~
       \mbox{and}~\partial_{R_2}V^{(0)}=0~\mbox{with}~R_1,R_2 \neq 0$\\[0.2cm]
       Solving with respect to $R_1,R_2$ we get
       $R_1^2 = (m_2^2 - m_1^2)/g^2-(m_1^2+m_2^2)/g_2^2$, and
       $R_2^2 = (m_1^2 - m_2^2)/g^2-(m_1^2+m_2^2)/g_2^2$.
       However, stability along $R_1=\pm R_2,~\theta=\phi=0$ direction
       dictates positiveness of the $m_1^2+m_2^2$ combination
       $(m_1^2+m_2^2 > 2|m_3^2|)$ so one of $R_1^2, R_2^2$ will be
       negative. Thus, no stationary solution exists or the potential
       is unbounded from below.
  \item[(II)]$\sin(\varphi+\theta)=0,~\partial_{R_1}V^{(0)}=0~\mbox{and}~
       \partial_{R_2}V^{(0)}=0~\mbox{with}~R_1,R_2 \neq 0$\\[0.2cm]
       In our conventions we always have $m_3^2<0$, so smaller values
       for $V^{(0)}$ are obtained for $\cos(\varphi+\theta)>0$. So we
       should take here $\cos(\varphi+\theta)=1$ (\ie~$\theta=-\varphi$).
       The rest of the extremum conditions will become
       \be
       \begin{array}{c}
       m_1^2 R_1  + m_3^2 R_2  + \hat{g}^2 R_1 (R_1^2  - R_2^2 )=0\\
       m_2^2 R_2  + m_3^2 R_1  - \hat{g}^2 R_2 (R_1^2  - R_2^2 ) = 0 \\
       \end{array} \label{stc2}
       \ee
       where $\hat{g}^2=(g^2  + g_2^2)/4$.
       Solutions for the above system are well known \cite{moul}. If we define
       $t=R_2/R_1$ (obviously $t=1$ does not satisfy Eq.~(\ref{stc2})),
       the solution in the ordinary case of radiative $SU(2)$ breaking
       $(m_1^2  + m_2^2  \pm 2m_3^2  >
       0,~m_1^2 >m_2^2,~m_1^2 m_2^2 - m_3^4 <0)$ is:
       \bea
       t &=&  \frac{- m_1^2  - m_2^2  - \sqrt {(m_1^2  + m_2^2  + 2m_3^2 )
       (m_1^2  + m_2^2  - 2m_3^2 )}}{2 m_3^2} \nonumber\\
       R_1 &=& \sqrt {\frac{{m_1^2  - m_2^2 t^2 }}{{\hat g^2 (t^4  - 1)}}}
       \ena
       and the global minimum configuration is:
       \be \left(
       \begin{array}{c}
       H_1 \\ H_2 \\ h_1 \\ h_2 \\
       \end{array}\right)_{min}= \left(\begin{array}{c}
       R_1 \cos\varphi   \\  t R_1 \cos\varphi  \\
       R_1 \sin\varphi   \\ -t R_1 \sin\varphi \\
       \end{array}\right)~~~~~~~~~~~\varphi \mbox{ arbitrary.}
       \ee
       It is trivial to show that the value at the minimum for this
       configuration coincides with the physical vacuum one of Eq.~(\ref{physv}).
\end{description}
{\centering \section{\label{th_ccb}Tree level constraints for
CCB-E\lowercase{b} direction}}%
Following closely the notation of \cite{sher,casas}, we express all fields in
Eq.~(\ref{colbr}) in terms of $|H_1|$, so that $|\sle_1| = \alpha
|H_1|,~|\sEC_1|=\beta |H_1|,~|H_2|=\gamma |H_1|$. Since the trilinear terms of
our example have small coupling $(|Y_e|^2 \ll 1)$, $D$-terms should be
suppressed. This implies that $\alpha=\beta$ and $\alpha^2 + \gamma^2 =1$. To
proceed further, two cases should be examined: \vspace{0.2cm}
\begin{description}
  \item[\textbullet] $\mathrm{sign}(A_e)=-\mathrm{sign}(B)$ with $m_3^2=\mu B$,
    $\alpha=\sqrt{1-\gamma^2}$ and $0<|\gamma|<1$.\\[0.2cm]
    Then, all three trilinear terms can be made negative simultaneously and
    the tree level potential becomes
    \be
    V_{\mathit{CCB-Eb}}^{(0)} = |Y_e|^2 \alpha^4 F(\alpha)
    |H_1|^4 - 2 Y_e \hat{A} \alpha^2 |H_1|^3 + \hat{m}^2 |H_1|^2 \label{beta1}
    \ee
    where
    $F(\alpha) = 1 + 2/\alpha^2$, $\hat{A} = |A_e|+ |\mu| \gamma$ and
    $\hat{m}^2 = m_1^2 + (m_{\sLE_1}^2 +
    m_{\sEC_1}^2) \alpha^2 + m_2^2 \gamma^2 - 2 |m_3^2|\gamma$. Differentiating
    with respect to $|H_1|$ for fixed values of $\gamma$, we find besides the
    trivial extremum the following local minimum for Eq.~(\ref{beta1})
    \be
    |H_1|_{\mathit{ext}} = \frac{3\hat{A}}{4Y_e \alpha^2 F(\alpha)}
    \left[ 1+\sqrt{1-\frac{8\hat{m}^2 F(\alpha)}{9\hat{A}^2}} \right]
    \ee
    Note that the typical vevs are of order $|A_e|/Y_e$. The corresponding
    value of the potential is
    \be
    V_{\mathit{ext}}=-\frac{1}{2}\alpha^2 |H_1|_{\mathit{ext}}^2 \left[ Y_e
    \hat{A}|H_1|_{\mathit{ext}} - \frac{\hat{m}^2}{\alpha^2} \right]
    \ee
    and the constraint to avoid a deeper configuration than the physical vacuum
    Eq.~(\ref{physv}) of the theory reads $\hat{A}^2 \leq F \hat{m}^2$ \ie
    \be
    (|A_e|+ |\mu| \gamma)^2 \leq \left( 1 + \frac{2}{\alpha^2} \right) \left(
    m_1^2 + (m_{\sLE_1}^2 +  m_{\sEC_1}^2) \alpha^2 + m_2^2 \gamma^2 -
    2 |m_3^2|\gamma \right)
    \ee
  \item[\textbullet] $\mathrm{sign}(A_e)=\mathrm{sign}(B)$ with
    $\alpha=\sqrt{1-\gamma^2}$ and $0<|\gamma|<1$. \\[0.2cm]
    Similarly the relevant constraints are
    \bea
    (|A_e|- |\mu| \gamma)^2 &\leq& \left( 1 + \frac{2}{\alpha^2} \right) \left(
    m_1^2 + (m_{\sLE_1}^2 +  m_{\sEC_1}^2) \alpha^2 + m_2^2 \gamma^2 -
    2 |m_3^2|\gamma \right)   \\
    (|A_e|+ |\mu| \gamma)^2 &\leq& \left( 1 + \frac{2}{\alpha^2} \right) \left(
    m_1^2 + (m_{\sLE_1}^2 +  m_{\sEC_1}^2) \alpha^2 + m_2^2 \gamma^2 +
    2 |m_3^2|\gamma \right)
    \ena
\end{description}
{\centering \section{\label{ufb1}Field dependent mass matrix elements for
UFB-1}}%
We cite here all the necessary mass matrix elements used in the definition of
the one loop effective potential. Let $\psi^r,~r=1,2,3$ stands for a Yukawa
($Y$) or a trilinear soft coupling ($A$). The following notation is used
throughout $\psi_u^r \equiv (\psi_u,~\psi_c,~\psi_t)$ and $\psi_d^r \equiv
(\psi_d,~\psi_s,~\psi_b)$.
\\[2mm] \textbf{\textbullet~Gauge Bosons ($\Omega=C=1$)}%
\bedm
  \lmv = \frac{1}{2} \lpa W_{\mu}^+~W_{\mu}^-~A_{\mu}~Z_{\mu} \rpa
  [\msv] \lpa  \begin{array}{c} W_{\mu}^- \\ W_{\mu}^+ \\  A_{\mu} \\
  Z_{\mu} \end{array}    \rpa,~~~~~~~~[\msv] =  \lpa
  \begin{array}{cccc}
  V_1   &   0 & V_2   & V_3 \\
    0   & V_1 & V_2^* & V_3^* \\
  V_2^* & V_2 & V_4   & V_5 \\
  V_3^* & V_3 & V_5   & V_6
  \end{array}   \rpa
\endm
\begin{eqnarray*}
  V_1 &=& \frac{g^2_2}{2} \lpa |H_1|^2 + |H_2|^2 + |h_1|^2 + |h_2|^2
  \rpa   \qquad   V_2 = -\frac{e g_2}{\sqrt{2}}
  \lpa  \cH_1 h_1 - H_2 \ch{2} \rpa       \\
  V_3 &=& \frac{e g}{\sqrt{2}} \lpa  \cH_1 h_1 - H_2 \ch{2} \rpa
      \qquad\qquad\quad\,
  V_4 = 2 e^2 \lpa |h_1|^2 + |h_2|^2 \rpa    \\
  V_5 &=&  -g g_2 \frac{g^2 - g_2^2}{g^2+g_2^2} \lpa  |h_1|^2 + |h_2|^2
         \rpa      \\
  V_6 &=&  \frac{g^2 + g_2^2}{2} \left[
     |H_1|^2 + |H_2|^2 + \lpa \frac{g^2 - g_2^2}{g^2+g_2^2} \rpa^2
     \lpa |h_1|^2 + |h_2|^2 \rpa
  \right]
\end{eqnarray*}
where $e=g g_2 / \sqrt{g^2+g_2^2}.$
\\[2mm] \textbf{\textbullet~Leptons $(\Omega=C=1)$}%
 \bedm
  M_e^2    = |Y_e|^2 (|H_1|^2 + |h_1|^2)~\qquad~
  M_\mu^2  = |Y_\mu|^2 (|H_1|^2 + |h_1|^2)~\qquad~
  M_\tau^2 = |Y_\tau|^2 (|H_1|^2 + |h_1|^2)
\endm
\textbf{\textbullet~Quarks ($\Omega=2,~C=3$)}%
\bedm
  \mathcal{L}_L = - \sum_{r,a=1}^3 \lpa u_{ra}^c~d_{ra}^c \rpa
  [\mathcal{M}_Q]_r \lpa \begin{array}{c} u_r^a \\  d_r^a
  \end{array}   \rpa~~\mbox{where}~~[\mathcal{M}_Q]_r = \lpa
  \begin{array}{cc}
  -Y_u^r H_2  & Y_u^r h_2 \\
  -Y_d^r h_1 & Y_d^r H_1
  \end{array} \rpa
\endm
The necessary eigenvalues should be computed from $\mathcal{M}_Q
\mathcal{M}_Q^{\dag}$.
\\[2mm] \textbf{\textbullet~Higgsinos ($\Omega=C=1$)}
\bedm
  \mathcal{L}_{\tilde{H}} = -\frac{1}{2} \chi^T
  [\mathcal{M}_{\tilde{H}}] \chi ~~\mbox{where}~~~\chi^T = \lpa
\tilde{H}^0_1~\tilde{H}_1^-~\tilde{H}_2^+~\tilde{H}^0_2~\tilde{B}~
\tilde{W}^+~\tilde{W}^-~\tilde{W}^{(3)}\rpa \mbox{ and}
\endm
$~
[\mathcal{M}_{\tilde{H}}] = \lpa \begin{array}{cc} S_1   & S_2^* \\
S_2^\dag & S_4 \end{array} \rpa \mbox{with}~~~~S_1 = \lpa
\begin{array}{rccr}
  0 & 0 & 0 & -\mu \\
  0 & 0 & \mu & 0 \\
  0 & \mu & 0 & 0 \\
  -\mu & 0 & 0 & 0
  \end{array}   \rpa,$
\bedm
  S_2 = \lpa
  \begin{array}{cccc}
  \frac{g}{\sqrt{2}} H_1   & 0         & -g_2 h_1& -\frac{g_2}{\sqrt{2}} H_1 \\
  \frac{g}{\sqrt{2}} h_1  & -g_2 H_1 & 0         & \frac{g_2}{\sqrt{2}} h_1 \\
  -\frac{g}{\sqrt{2}} h_2 & 0         & -g_2 H_2 & -\frac{g_2}{\sqrt{2}} h_2\\
  -\frac{g}{\sqrt{2}} H_2  & -g_2 h_2& 0         & \frac{g_2}{\sqrt{2}} H_2
  \end{array}   \rpa,~~~
  S_4 = \lpa \begin{array}{cccc}
  M_1 & 0 & 0 & 0 \\
  0 & 0 & M_2 & 0 \\
  0 & M_2 & 0 & 0 \\
  0 & 0 & 0 & M_2
\end{array}   \rpa.
\endm
\\[2mm]%
\textbf{\textbullet~Sleptons $(\sLE^T \equiv (~ \sne~~\sle ~)~\Omega=2, ~C=1)$}
\bedm
  \mathcal{L}_{\tilde{L}} =
  - \sum_{r=1}^3 \lpa \sne_r~\sle_r~\csEC_r  \rpa
  [\mathcal{M}_{\tilde{L}}^2]_r \lpa \begin{array}{c}
  \sne_r^* \\    \sle_r^* \\     \sEC_r
  \end{array}   \rpa~~\mbox{where}~~[\mathcal{M}_{\tilde{L}}^2]_r = \lpa
  \begin{array}{ccc}
  B_1^r & B_2^r & A_1^r \\ B_2^{r*} & B_3^r & A_2^r \\ A_1^{r*} & A_2^{r*} & A_3^r
  \end{array}   \rpa.
\endm
We give below the relevant entries
\begin{eqnarray*}
  A_1^r &=& Y_e^r (\mu \ch{2} - A_e^r h_1)   \qquad\qquad\qquad
  A_2^r = Y_e^r (\mu \cH_2 + A_e^r H_1)   \\
  A_3^r &=& m_{\sEC_r}^2 + |Y_e^r|^2 \lpa |H_1|^2 + |h_1|^2 \rpa -
     \frac{g^2}{2} \lpa |H_1|^2 + |h_1|^2 - |H_2|^2 - |h_2|^2 \rpa   \\
  B_1^r &=& m_{\sLE_r}^2 + \magf{Y_e^r h_1} +
     \frac{g^2+g_2^2}{4} \lpa |H_1|^2 - |H_2|^2 \rpa +
     \frac{g^2-g_2^2}{4} \lpa |h_1|^2 - |h_2|^2 \rpa  \\
  B_2^r &=& \lpa \frac{g_2^2}{2} - |Y_e^r|^2 \rpa \cH_1 h_1 +
     \frac{g_2^2}{2} H_2 \ch{2}   \\
  B_3^r &=&  m_{\sLE_r}^2 + \magf{Y_e^r H_1} +
     \frac{g^2-g_2^2}{4} \lpa |H_1|^2 - |H_2|^2 \rpa +
     \frac{g^2+g_2^2}{4} \lpa |h_1|^2 - |h_2|^2 \rpa   \\
\end{eqnarray*}
\textbf{\textbullet~Squarks $(\sQU^T \equiv (~ \squ~~\sqd ~)~\Omega=2,~C=3)$}
\bedm
  \mathcal{L}_{\tilde{Q}} = -\sum_{a,r=1}^3
  \lpa \squ_r^a~\sqd_r^a~\csUC_{ra}~\csDC_{ra} \rpa
  [M_{\tilde{Q}}^2]_r \lpa
  \begin{array}{c}
  \squ_r^{a*} \\ \sqd_r^{a*} \\ \sUC_{ra} \\ \sDC_{ra}
  \end{array}   \rpa~~\mbox{where}~~[M_{\tilde{Q}}^2]_r= \lpa
  \begin{array}{cccc}
     \Lambda_1^r    & \Lambda_2^r    & K_5^r    & K_7^r \\
     \Lambda_2^{r*} & \Lambda_3^r    & K_6^r    & K_8^r \\
     K_5^{r*}       & K_6^{r*}       & K_1^r    & K_2^r \\
     K_7^{r*}       & K_8^{r*}       & K_2^{r*} & K_3^r
  \end{array}    \rpa
\endm
where
\begin{eqnarray*}
   \Lambda_1^r &=& m_{\sQU_r}^2 + \magf{Y_u^r H_2} + \magf{Y_d^r h_1}
   - G^- \lpa |H_1|^2 - |H_2|^2 \rpa
   - G^+ \lpa |h_1|^2 - |h_2|^2 \rpa   \\
 \Lambda_2^r &=& \lpa \frac{g_2^2}{2} - |Y_d^r|^2 \rpa \cH_1 h_1
    + \lpa \frac{g_2^2}{2} - |Y_u^r|^2 \rpa H_2 \ch{2}  \\
 \Lambda_3^r &=& m_{\sQU_r}^2 + \magf{Y_d^r H_1} + \magf{Y_u^r h_2}
   - G^+ \lpa |H_1|^2 - |H_2|^2 \rpa
   - G^- \lpa |h_1|^2 - |h_2|^2 \rpa   \\
 K_1^r &=& m_{\sUC_r}^2 + |Y_u^r|^2 \lpa |H_2|^2 + |h_2|^2 \rpa
   + \frac{g^2}{3} \lpa |H_1|^2 + |h_1|^2 - |H_2|^2 - |h_2|^2 \rpa  \\
 K_2^r &=& Y_u^r Y_d^r \lpa H_1 \ch{2} + \cH_2 h_1 \rpa    \\
 K_3^r &=&  m_{\sDC_r}^2 + |Y_d^r|^2 \lpa |H_1|^2 + |h_1|^2 \rpa
   - \frac{g^2}{6} \lpa |H_1|^2 + |h_1|^2 - |H_2|^2 - |h_2|^2 \rpa
\end{eqnarray*}
and by definition $G^+ = (g^2+3g_2^2)/12~~~~~G^- = (g^2-3g_2^2)/12.$ Besides
\bedm
  \begin{array}{cccrc}
  K_5^r &=& -Y_u^r (\mu \cH_1 + A_u^r H_2) \qquad\qquad
   & K_6^r =& -Y_u^r (\mu \ch{1} - A_u^r h_2)  \\
  K_7^r &=& Y_d^r (\mu \ch{2} - A_d^r h_1) \qquad\qquad
   & K_8^r =& Y_d^r (\mu \cH_2 + A_d^r H_1)
  \end{array}
\endm
\textbf{\textbullet~Higgs $(\Omega=1,~C=1)$}%
\bedm
  \mathcal{L}_{H} = -\frac{1}{2} \phi^T [\mathcal{M}_H^2]
  \phi^*~~\mbox{where}~~ [\mathcal{M}_H^2] = \lpa \begin{array}{cc}
  T_1 & T_2^\dag \\ T_2 & T_1^* \end{array} \rpa
\endm
with \bedm
  T_1 = \lpa \begin{array}{cccc}
  R_1   & R_2   & R_4   & R_5 \\
  R_2^* & R_3   & R_6   & R_7 \\
  R_4^* & R_6^* & R_8   & R_9 \\
  R_5^* & R_7^* & R_9^* & R_{10}
\end{array}  \rpa, \qquad\qquad T_2 = \lpa \begin{array}{cccc}
  Q_1 & Q_2 & Q_4 & Q_5 \\
  Q_2 & Q_3 & Q_6 & Q_7 \\
  Q_4 & Q_6 & Q_8 & Q_9 \\
  Q_5 & Q_7 & Q_9 & Q_{10}
\end{array}  \rpa
\endm
and $\phi^T = \lpa \varphi^T~\varphi^{T*} \rpa,~~\varphi^T = \lpa
H_1^0~H_1^-~H_2^+~H_2^0 \rpa.$ Additionally
\begin{eqnarray*}%
  R_1 &=& m_1^2 + \frac{g^2+g_2^2}{4}
    \lpa 2 |H_1|^2 + |h_1|^2 - |H_2|^2 - |h_2|^2 \rpa
    + \frac{g_2^2}{2} |h_2|^2       \\
  R_2 &=& \frac{g^2+g_2^2}{4} \cH_1 h_1
    + \frac{g_2^2}{2} H_2 \ch{2}    \\
  R_3 &=& m_1^2 + \frac{g^2+g_2^2}{4}
    \lpa |H_1|^2 + 2|h_1|^2 - |H_2|^2 - |h_2|^2 \rpa
    + \frac{g_2^2}{2} |H_2|^2       \\
  R_4 &=& \frac{g_2^2}{2} H_2 \ch{1} - \frac{g^2-g_2^2}{4} \cH_1
    h_2 \qquad\qquad  R_5 = -\frac{g^2+g_2^2}{4} \cH_1 H_2  \\
  R_6 &=& -\frac{g^2+g_2^2}{4} \ch{1} h_2  \qquad\qquad~~~~~~~~~~~~\,
    R_7 = -\frac{g^2-g_2^2}{4} H_2 \ch{1} + \frac{g_2^2}{2} \cH_1
    h_2   \\
  R_8 &=& m_2^2 + \frac{g^2+g_2^2}{4}
    \lpa |H_2|^2 + 2|h_2|^2 - |H_1|^2 - |h_1|^2 \rpa
    + \frac{g_2^2}{2} |H_1|^2          \\
  R_9 &=& \frac{g_2^2}{2} \cH_1 h_1
  + \frac{g^2+g_2^2}{4} H_2 \ch{2}     \\
  R_{10} &=& m_2^2 + \frac{g^2+g_2^2}{4}
    \lpa 2|H_2|^2 + |h_2|^2 - |H_1|^2 - |h_1|^2 \rpa
    + \frac{g_2^2}{2} |h_1|^2
\end{eqnarray*}
and the matrix element for $T_2$ are%
\bedm
\begin{array}{llll} ~~~~Q_1 =& \displaystyle \frac{g^2+g_2^2}{4}
  (H_1)^2 \qquad &Q_2 =&
  \displaystyle \frac{g^2+g_2^2}{4} H_1 h_1  \\[3mm]
  ~~~~Q_3 =& \displaystyle \frac{g^2+g_2^2}{4} (h_1)^2 \qquad &Q_4 =&
  -\displaystyle \frac{g^2-g_2^2}{4}
  H_1 h_2  \\[3mm]
  ~~~~Q_5 =&\displaystyle m_3^2 - \frac{g^2+g_2^2}{4} H_1 H_2 +
  \frac{g_2^2}{2} h_1 h_2 \qquad &Q_6 =&\displaystyle -m_3^2 -
  \frac{g^2+g_2^2}{4} h_1 h_2 + \frac{g_2^2}{2} H_1 H_2 \\[3mm]
  ~~~~Q_7 =&\displaystyle -\frac{g^2-g_2^2}{4} H_2 h_1 &Q_8 =&
  \displaystyle \frac{g^2+g_2^2}{4} (h_2)^2  \\[3mm]
  ~~~~Q_9 =&\displaystyle \frac{g^2+g_2^2}{4} H_2 h_2 &Q_{10} =&
  \displaystyle \frac{g^2+g_2^2}{4} (H_2)^2 \\[3mm]
\end{array}
\endm
{\centering \section{\label{ufb3b}Field dependent mass matrix elements for
UFB-3\lowercase{b}}}
 Here we present the mass matrix elements needed in the one
loop effective potential expression. We use the same notation as in Appendix
\ref{ufb1}.
\\[2mm]
\textbf{\textbullet~Gauge Bosons $(\Omega_{W^\pm}=2,~\Omega_{A,Z}=1,~C=1)$}
\bedm
  \lmv = M_{W^{\pm}}^2 W_{\mu}^+ W^{-\mu} + \frac{1}{2} \lpa A_{\mu}~
  Z_{\mu} \rpa [M^2_{(A,Z)}] \lpa \begin{array}{c} A^{\mu} \\ Z^{\mu}
  \end{array} \rpa, \qquad  [M^2_{(A,Z)}]=\lpa \begin{array}{cc}
  V_2 & V_3 \\ V_3 & V_4 \end{array} \rpa
\endm
\begin{eqnarray*}
  M_{W^{\pm}}^2 &=& \frac{g_2^2}{2}\lpa |H_2|^2 + |\sne_2|^2
  + |\sle_3|^2 \rpa      \\
  V_2 &=& 2 e^2 \lpa |\sle_3|^2 + |\sEC_3|^2 \rpa ~~~~~~~~~~
  V_3 = g g_2 \left[ |\sle_3|^2 - 2 \snw{2} \lpa
     |\sEC_3|^2 + |\sle_3|^2 \rpa \right]      \\
  V_4 &=& \frac{g^2+g_2^2}{2}\left[ |H_2|^2 + |\sne_2|^2
     + (1 - 2 \snw{2})^2 |\sle_3|^2 + 4 \snw{4} |\sEC_3|^2
  \right]
\end{eqnarray*}
where $e=g g_2 / \sqrt{g^2+g_2^2}$ and $\snw{1}=e/g_2.$
\\[2mm]\textbf{\textbullet~Quarks $(\Omega=2,~C=3)$}%
\bedm
  M_u^2 = |Y_u H_2|^2~\qquad~M_c^2 = |Y_c H_2|^2~\qquad~ M_t^2
  = |Y_t H_2|^2
\endm
\textbf{\textbullet~Leptons - Higgsinos $(\Omega=1,~C=1)$}%
\bedm
  \mathcal{L}_{(L,\tilde{H})} = -\frac{1}{2} \psi_1^T [M_1^f] \psi_1 -
  \frac{1}{2} \psi_2^T [M_2^f] \psi_2
\endm
where $\psi_1^T = \lpa
\nu_{\mu}~\tau~\tau^c~\tilde{H}_2^0~\tilde{H}_1^0~\tilde{B}~ \tilde{W}^{(3)}
\rpa $ and $\psi_2^T = \lpa
\nu_{\tau}~\mu~\mu^c~\tilde{H}_2^+~\tilde{H}_1^-~\tilde{W}^+~\tilde{W}^-
\rpa.$ Besides%
\bedm
  [M_1^f] = \lpa \begin{array}{ccccccc}
  0 & 0 & 0 & 0 & 0 & \frac{g}{\sqrt{2}}\csne_2 & -\frac{g_2}{\sqrt{2}}\csne_2 \\
  0 & 0 & 0 & 0 & Y_{\tau} \sEC_3 & \frac{g}{\sqrt{2}}\csle_3 &
  \frac{g_2}{\sqrt{2}}\csle_3 \\
  0 & 0 & 0 & 0 & Y_{\tau} \sle_3 & -\sqrt{2} g \csEC_3 & 0               \\
  0 & 0 & 0 & 0 & \mu & -\frac{g}{\sqrt{2}}\cH_2 & \frac{g_2}{\sqrt{2}}\cH_2  \\
  0 & Y_{\tau} \sEC_3 & Y_{\tau} \sle_3 & \mu & 0 & 0 & 0 \\
  \frac{g}{\sqrt{2}}\csne_2 & \frac{g}{\sqrt{2}}\csle_3 & -\sqrt{2} g \csEC_3 &
  -\frac{g}{\sqrt{2}}\cH_2 & 0 & -M_1 & 0 \\
  -\frac{g_2}{\sqrt{2}}\csne_2 & \frac{g_2}{\sqrt{2}}\csle_3  & 0 &
  \frac{g_2}{\sqrt{2}}\cH_2   & 0 & 0 & -M_2
  \end{array}  \rpa
\endm
\bedm
  [M_2^f] = \lpa \begin{array}{ccccccc}
  0 & 0 & 0 & 0 & -Y_{\tau} \sEC_3 & 0 & -g_2 \csle_3 \\
  0 & 0 & 0 & 0 & 0 & -g_2 \csne_2 & 0 \\
  0 & 0 & 0 & 0 & -Y_{\mu} \sne_2 & 0 & 0 \\
  0 & 0 & 0 & 0 & -\mu & 0 & -g_2 \cH_2 \\
  -Y_{\tau} \sEC_3 & 0 & -Y_{\mu} \sne_2 & -\mu & 0 & 0 & 0 \\
  0 & -g_2 \csne_2 & 0 & 0 & 0 & 0 & -M_2 \\
  -g_2 \csle_3 & 0 & 0 & -g_2 \cH_2 & 0 & -M_2 & 0
  \end{array}  \rpa
\endm
\\[2mm]\textbf{\textbullet~Squarks $(\sQU^T \equiv (~ \squ~~\sqd
~),~\Omega=2,~C=3)$} \bedm
  \mathcal{L}_{\tilde{Q}} = -\sum_{r,a=1}^3 \lpa \sUC_{ra}~\squ_r^{a*}
  \rpa [M_1^2] \lpa \begin{array}{c} \csUC_{ra} \\ \squ_r^a
  \end{array} \rpa - \sum_{r,a=1}^3 \lpa \sDC_{ra}~\sqd_r^{a*} \rpa
  [M_2^2] \lpa \begin{array}{c} \csDC_{ra} \\ \sqd_r^a
  \end{array} \rpa
\endm
where $[M_1^2]= \lpa \begin{array}{cc} K_1^r & K_4^r \\  K_4^{r*} & \Lambda_1^r
\end{array}  \rpa$ and $[M_2^2]= \lpa \begin{array}{cc} K_2^r & K_5^r \\
K_5^{r*} & \Lambda_2^r \end{array}  \rpa.$ The relevant quantities are

\begin{eqnarray*}
  K_1^r &=& m_{\sUC_r}^2 + \magf{Y_u^r H_2} -\frac{g^2}{3} \lpa
    2|\sEC_3|^2 + |H_2|^2 - |\sne_2|^2 - |\sle_3|^2   \rpa         \\
  K_4^r &=& -Y_u^r A_u^r H_2     \\
  \Lambda_1^r &=& m_{\sQU_r}^2 + \magf{Y_u^r H_2}
  + \frac{g^2}{6} \lpa |\sEC_3|^2 - |\sle_3|^2 \rpa
  + \frac{g^2 - 3g_2^2}{12} \lpa |H_2|^2 + |\sle_3|^2 - |\sne_2|^2 \rpa
\end{eqnarray*}%
\begin{eqnarray*}
  K_2^r &=& m_{\sDC_r}^2 + \frac{g^2}{6} \lpa 2|\sEC_3|^2 + |H_2|^2
    - |\sne_2|^2 - |\sle_3|^2 \rpa \\
  K_5^r &=& Y_d^r (\mu \cH_2 + Y_{\tau} \csle_3 \csEC_3)\\
  \Lambda_2^r &=& m_{\sQU_r}^2
  + \frac{g^2}{6} \lpa |\sEC_3|^2 - |\sle_3|^2 \rpa
  + \frac{g^2 + 3g_2^2}{12} \lpa |H_2|^2 + |\sle_3|^2 - |\sne_2|^2 \rpa
\end{eqnarray*}
\textbf{\textbullet~Sleptons - Higgses $(\sLE^T \equiv (~ \sne~~\sle ~))$}
\bedm
  \mathcal{L}_{(\tilde{L},H)} = -M^2 |\sne_1|^2 - \lpa
  \sEC_1~\sle_1^* \rpa [M_{s_1}^2] \lpa \begin{array}{c}
  \csEC_1 \\ \sle_1 \end{array}  \rpa  - \chi^T [M_{s_2}^2] \chi^*
  -\frac{1}{2} \phi^T [M_{s_3}^2] \phi^*
\endm
where $\chi^T = \lpa H_2^+~\sEC_2~\sne_3~H_1^{-*}~\sle_2^* \rpa $ and $\phi^T =
\lpa \xi^T~\xi^{T*} \rpa,~\xi^T = \lpa H_1^0~H_2^0~\sEC_3~
\sne_2~\sle_3 \rpa.$%
\bedm
  M^2 = m_{\sLE_1}^2 + \frac{g^2}{2} \lpa |\sle_3|^2 - |\sEC_3|^2 \rpa
  + \frac{g^2 + g_2^2}{4} \lpa |\sne_2|^2 - |\sle_3|^2
  - |H_2|^2 \rpa~~~~~~(\Omega=2,~C=1).
\endm
On the other hand $[M_{s_1}^2] = \lpa \begin{array}{cc} A_1 & A_2 \\
A_2^* & A_3  \end{array} \rpa~(\Omega=2,~C=1)$ with
\begin{eqnarray*}
  A_1 &=& m_{\sEC_1}^2 + \frac{g^2}{2} \lpa 2|\sEC_3|^2 + |H_2|^2
     - |\sne_2|^2 - |\sle_3|^2 \rpa~~~~~~~~
  A_2 = Y_e (\mu \cH_2 + Y_{\tau} \csle_3 \csEC_3)  \\
  A_3 &=& m_{\sLE_1}^2 + \frac{g^2}{2} \lpa |\sle_3|^2 - |\sEC_3|^2 \rpa
  + \frac{g^2 - g_2^2}{4} \lpa |\sne_2|^2 - |\sle_3|^2
  - |H_2|^2 \rpa.
\end{eqnarray*}

Similarly $[M_{s_2}^2] = \lpa \begin{array}{ccccc}
    B_1      & B_2   & B_3      & -m_3^2 & B_4    \\
    B_2^*    & B_5   & B_6      & B_7    & B_8    \\
    B_3^*    & B_6^* & B_9      & B_{10} & B_{11} \\
    -m_3^2   & B_7^* & B_{10}^* & B_{12} & 0      \\
    B_4^*    & B_8^* & B_{11}^* & 0      & B_{13}
\end{array} \rpa~(\Omega=2,~C=1)$ with
\begin{eqnarray*}
  B_1 &=& m_2^2 + \frac{g^2}{2} \lpa |\sEC_3|^2 - |\sne_2|^2\rpa
    + \frac{g^2+g_2^2}{4} \lpa |\sne_2|^2 + |H_2|^2 - |\sle_3|^2 \rpa     \\
  B_2 &=& \mu Y_{\mu} \csne_2  ~~~~~~~~~~~~~~\,
  B_3 = \mu Y_{\tau} \csEC_3 + \frac{g_2^2}{2} \cH_2 \sle_3   ~~~~~~~~~
  B_4 = \frac{g_2^2}{2} \csne_2 \cH_2    \\
  B_5 &=& m_{\sEC_2}^2 + \magf{Y_{\mu} \sne_2} + \frac{g^2}{2}
    \lpa 2|\sEC_3|^2 + |H_2|^2 - |\sne_2|^2 - |\sle_3|^2 \rpa  \\
    B_6 &=& Y_{\mu} Y_{\tau} \sne_2 \csEC_3    ~~~~~~~~~
  B_7 = -Y_{\mu} A_{\mu} \sne_2              ~~~~~~~~~~~~~~~~~~~~
  B_8 = Y_{\mu} (\mu \cH_2 + Y_{\tau} \csle_3 \csEC_3)
\end{eqnarray*}
\begin{eqnarray*}
  B_9 &=& m_{\sLE_3}^2 + \lpa \magf{Y_{\tau}} - \frac{g^2}{2} \rpa |\sEC_3|^2 +
    \frac{g^2+g_2^2}{4} \lpa |\sne_2|^2 + |\sle_3|^2  - |H_2|^2  \rpa     \\
  B_{10} &=& -Y_{\tau} A_{\tau} \sEC_3    ~~~~~~~~~~
  B_{11} =  \frac{g_2^2}{2} \csne_2 \csle_3  \\
  B_{12} &=&  m_1^2 + \magf{Y_{\mu} \sne_2} + \magf{Y_{\tau} \sEC_3}
  + \frac{g^2}{2} \lpa |\sle_3|^2 - |\sEC_3|^2 \rpa
    + \frac{g^2 - g_2^2}{4} \lpa |\sne_2|^2 - |\sle_3|^2 - |H_2|^2 \rpa  \\
  B_{13} &=& m_{\sLE_2}^2 + \frac{g^2+g_2^2}{4} \lpa |\sne_2|^2 + |\sle_3|^2
    + |H_2|^2 \rpa - \frac{g^2}{2} \lpa |H_2|^2 + |\sEC_3|^2 \rpa
\end{eqnarray*}
and $[M_{s_3}^2] = \lpa \begin{array}{cc} D_1 & D_2^\dag \\ D_2 & D_1^*
\end{array} \rpa~(\Omega=C=1)$ with
\bedm
  D_1 = \lpa
  \begin{array}{ccccc}
      R_1 & 0   & 0     & 0        & 0      \\
      0 & R_2   & R_3   & R_4      & R_5    \\
      0 & R_3^* & R_6   & R_7      & R_8    \\
      0 & R_4^* & R_7^* & R_9      & R_{10} \\
      0 & R_5^* & R_8^* & R_{10}^* & R_{11}
  \end{array} \rpa, \qquad ~~~\qquad  D_2 = \lpa
  \begin{array}{ccccc}
     0     & m_3^2 & Q_1 & 0      & Q_2    \\
     m_3^2 & Q_3   & Q_4 & Q_5    & Q_6    \\
     Q_1   & Q_4   & Q_7 & Q_8    & Q_9    \\
     0     & Q_5   & Q_8 & Q_{10} & Q_{11} \\
     Q_2   & Q_6   & Q_9 & Q_{11} & Q_{12}
  \end{array}  \rpa
\endm%
\begin{eqnarray*}
  R_1 &=& m_1^2 + \magf{Y_{\tau}} \lpa |\sEC_3|^2 + |\sle_3|^2 \rpa
     + \frac{g^2}{2} \lpa |\sle_3|^2 - |\sEC_3|^2 \rpa
     + \frac{g^2+g_2^2}{4} \lpa |\sne_2|^2 - |\sle_3|^2 - |H_2|^2 \rpa \\
  R_2 &=& m_2^2
     + \frac{g^2}{2} \lpa |\sEC_3|^2 - |\sle_3|^2 \rpa
     + \frac{g^2+g_2^2}{4} \lpa 2|H_2|^2 + |\sle_3|^2 - |\sne_2|^2 \rpa   \\
  R_3 &=& \mu Y_{\tau} \csle_3 + \frac{g^2}{2} \cH_2 \sEC_3   ~~~~~~~~
  R_4 = -\frac{g^2+g_2^2}{4} \cH_2 \sne_2   ~~~~~~~~
  R_5 = \mu Y_{\tau} \csEC_3 - \frac{g^2-g_2^2}{4} \cH_2 \sle_3    \\
  R_6 &=& m_{\sEC_3}^2 + \magf{Y_{\tau} \sle_3} +
     \frac{g^2}{2} \lpa 4|\sEC_3|^2 + |H_2|^2 - |\sne_2|^2
     - |\sle_3|^2 \rpa    \\
  R_7 &=& -\frac{g^2}{2} \csEC_3 \sne_2    ~~~~~~~~~~~~~~~~~
  R_8 = \lpa |Y_{\tau}|^2 - \frac{g^2}{2} \rpa \csEC_3 \sle_3   \\
  R_9 &=& m_{\sLE_2}^2 + \frac{g^2}{2} \lpa |\sle_3|^2 - |\sEC_3|^2 \rpa
     + \frac{g^2+g_2^2}{4} \lpa 2|\sne_2|^2 - |\sle_3|^2 - |H_2|^2 \rpa   \\
  R_{10} &=& \frac{g^2-g_2^2}{4} \csne_2 \sle_3    \\
  R_{11} &=& m_{\sLE_3}^2 + \magf{Y_{\tau} \sEC_3}
     + \frac{g^2}{2} \lpa 2|\sle_3|^2 - |\sEC_3|^2 \rpa
     + \frac{g^2-g_2^2}{4} \lpa |\sne_2|^2 - 2|\sle_3|^2 -|H_2|^2 \rpa
\end{eqnarray*}
while the corresponding elements for $D_2$ are \bedm
\begin{array}{llllll} Q_1 =& Y_{\tau} A_{\tau} \csle_3 \qquad\qquad &Q_2 =&
  \displaystyle Y_{\tau} A_{\tau} \csEC_3 \qquad & Q_3 =& \displaystyle
  \frac{g^2+g_2^2}{4} (H_2)^2 \\[3mm]
  Q_4 =&   \displaystyle \frac{g^2}{2} H_2 \sEC_3  &
  Q_5 =&\displaystyle -\frac{g^2+g_2^2}{4} H_2 \sne_2 \qquad\qquad &Q_6 =&
  \displaystyle -\frac{g^2-g_2^2}{4} \sle_3 H_2
\end{array}
\endm
\bedm
\begin{array}{llll}
  Q_7 =&\displaystyle (g \sEC_3)^2   &Q_8 =&
  \displaystyle -\frac{g^2}{2} \sEC_3 \sne_2 \\[3mm]
  Q_9 =&\displaystyle \mu Y_{\tau} H_2 +
     \lpa |Y_{\tau}|^2 - \frac{g^2}{2} \rpa \sle_3 \sEC_3 \qquad\qquad\qquad\qquad
    &Q_{10} =&  \displaystyle \frac{g^2+g_2^2}{4} (\sne_2)^2  \\[3mm]
  Q_{11} =& \displaystyle \frac{g^2-g_2^2}{4} \sne_2 \sle_3 &Q_{12} =&
  \displaystyle \frac{g^2+g_2^2}{4} (\sle_3)^2 \\[5mm]
\end{array}
\endm
{\centering \section{\label{ccbeb}Field dependent mass matrix elements for
CCB-E\lowercase{b}}}%
Using the notation of Appendix \ref{ufb1}, the mass matrix elements needed in
the one loop effective potential expression
are\\[2mm]%
\textbf{\textbullet~Gauge Bosons $(\Omega_{W^\pm}=2,~\Omega_{A,Z}=1,~C=1)$}
\bedm
  \lmv = M_{W^{\pm}}^2 W_{\mu}^+ W^{-\mu} + \frac{1}{2} \lpa A_{\mu}~
  Z_{\mu} \rpa [M^2_0] \lpa \begin{array}{c} A^{\mu} \\ Z^{\mu}
  \end{array} \rpa, \qquad  [M^2_0]=\lpa \begin{array}{cc}
  V_2 & V_3 \\ V_3 & V_4 \end{array} \rpa
\endm
\begin{eqnarray*}
  M_{W^{\pm}}^2 &=& \frac{g_2^2}{2}\lpa |H_1|^2 + |H_2|^2
  + |\sle_1|^2 \rpa     \\
  V_2 &=& 2 e^2 \lpa |\sle_1|^2 + |\sEC_1|^2 \rpa~~~~~~~~~~~~
  V_3 = g g_2 \left[ |\sle_1|^2 - 2 \snw{2}
  \lpa |\sle_1|^2 + |\sEC_1|^2 \rpa \right]      \\
  V_4 &=& \frac{g^2+g_2^2}{2}\left[ |H_1|^2 + |H_2|^2
  + (1 - 2 \snw{2})^2 |\sle_1|^2 + 4 \snw{4} |\sEC_1|^2
  \right]
\end{eqnarray*}
where $e=g g_2 / \sqrt{g^2+g_2^2}$ and $\snw{1}=e/g_2.$
\\[2mm]
\textbf{\textbullet~Quarks $(\Omega=2,~C=3)$} \bedm
\begin{array}{c}
  M_u^2 = |Y_u H_2|^2~\qquad~M_c^2 = |Y_c H_2|^2~\qquad~ M_t^2
  = |Y_t H_2|^2 \\[2mm]
  M_d^2 = |Y_d H_1|^2~\qquad~M_s^2 = |Y_s H_1|^2~\qquad~ M_b^2
  = |Y_b H_1|^2
\end{array}
\endm
\textbf{\textbullet~Heavy Leptons $(\Omega=2,~C=1)$}
\bedm%
 M_{\mu}^2 = |Y_\mu H_1|^2~\qquad~M_\tau^2 = |Y_\tau H_1|^2
\endm
\textbf{\textbullet~Light Leptons - Higgsinos $(\Omega=1,~C=1)$} \bedm
  \mathcal{L}_2 = -\frac{1}{2} \psi_1^T [M_1^f] \psi_1 -
  \frac{1}{2} \psi_2^T [M_2^f] \psi_2
\endm
where $\psi_1^T = \lpa e~e^c~\tilde{H}_1^0~\tilde{H}_2^0~\tilde{B}~
\tilde{W}^{(3)} \rpa $ and $\psi_2^T = \lpa
\nu_e~\tilde{H}_1^-~\tilde{H}_2^+~\tilde{W}^+~\tilde{W}^- \rpa.$ Also \bedm
  [M_1^f] = \lpa \begin{array}{cccccc}
  0  & Y_e H_1  & Y_e \sEC_1  & 0  & \frac{g}{\sqrt{2}}\csle_1  &
  \frac{g_2}{\sqrt{2}}\csle_1 \\
  Y_e H_1  & 0  & Y_e \sle_1  & 0  & -\sqrt{2} g \csEC_1  & 0  \\
  Y_e \sEC_1  & Y_e \sle_1  & 0   & \mu  & \frac{g}{\sqrt{2}} \cH_1  &
  -\frac{g_2}{\sqrt{2}}\cH_1  \\
  0  & 0  & \mu  &  0  & -\frac{g}{\sqrt{2}} \cH_2  &
  \frac{g_2}{\sqrt{2}}\cH_2   \\
  \frac{g}{\sqrt{2}}\csle_1  & -\sqrt{2} g \csEC_1 & \frac{g}{\sqrt{2}} \cH_1 &
  -\frac{g}{\sqrt{2}} \cH_2  & -M_1  & 0   \\
  \frac{g_2}{\sqrt{2}}\csle_1  & 0  & -\frac{g_2}{\sqrt{2}}\cH_1 &
  \frac{g_2}{\sqrt{2}}\cH_2   & 0  & -M_2
  \end{array}  \rpa
\endm
\bedm
  [M_2^f] = \lpa \begin{array}{ccccc}
  0             & -Y_e \sEC_1 & 0           & 0          & -g_2 \csle_1 \\
  -Y_e \sEC_1   & 0           & -\mu        & -g_2 \cH_1 & 0            \\
  0             & -\mu        & 0           & 0          & -g_2 \cH_2   \\
  0             & -g_2 \cH_1  & 0           & 0          & -M_2         \\
  -g_2 \csle_1  & 0           & -g_2 \cH_2  & -M_2       & 0
  \end{array}  \rpa
\endm
\textbf{\textbullet~Squarks $(\sQU^T \equiv (~ \squ~~\sqd ~),~\Omega=2,~C=3)$}
\bedm
  \mathcal{L}_{\tilde{Q}} = -\sum_{r,a=1}^3 \lpa \sUC_{ra}~\squ_r^{a*}
  \rpa [M_1^2] \lpa \begin{array}{c} \csUC_{ra} \\ \squ_r^a
  \end{array} \rpa - \sum_{r,a=1}^3 \lpa \sDC_{ra}~\sqd_r^{a*} \rpa
  [M_2^2] \lpa \begin{array}{c} \csDC_{ra} \\ \sqd_r^a
  \end{array} \rpa
\endm
where $[M_1^2]= \lpa \begin{array}{cc} \Gamma_1^r & \Gamma_2^r \\
\Gamma_2^{r*} & \Gamma_3^r \end{array}  \rpa$ and $[M_2^2]= \lpa
\begin{array}{cc} \Delta_1^r & \Delta_2^r \\  \Delta_2^{r*} & \Delta_3^r
\end{array} \rpa.$ The relevant quantities are
\begin{eqnarray*}
  \Gamma_1^r &=& m_{\sUC_r}^2 + \magf{Y_u^r H_2} - \frac{g^2}{3} \lpa
      2|\sEC_1|^2 + |H_2|^2  - |H_1|^2 - |\sle_1|^2 \rpa  \\
  \Gamma_2^r &=& -Y_u^r (\mu \cH_1 + A_u^r H_2)     \\
  \Gamma_3^r &=& m_{\sQU_r}^2 + \magf{Y_u^r H_2}
  + \frac{g^2}{6} \lpa |\sEC_1|^2 - |\sle_1|^2 \rpa
  + \frac{g^2 - 3g_2^2}{12} \lpa |H_2|^2 + |\sle_1|^2 - |H_1|^2 \rpa  \\
  \Delta_1^r &=& m_{\sDC_r}^2 + \magf{Y_d^r H_1}
  + \frac{g^2}{6} \lpa 2|\sEC_1|^2 + |H_2|^2 - |H_1|^2 - |\sle_1|^2 \rpa \\
  \Delta_2^r &=& Y_d^r (\mu \cH_2 + A_d^r H_1 + Y_e \csle_1 \csEC_1) \\
  \Delta_3^r &=& m_{\sQU_r}^2 + \magf{Y_d^r H_1}
  + \frac{g^2}{6} \lpa |\sEC_1|^2 - |\sle_1|^2 \rpa
  + \frac{g^2 + 3g_2^2}{12} \lpa |H_2|^2 + |\sle_1|^2 - |H_1|^2 \rpa
\end{eqnarray*}
\textbf{\textbullet~Sleptons - Higgses $(\sLE^T \equiv (~ \sne~~\sle ~))$}
\begin{eqnarray*}
  \mathcal{L}_{(\tilde{L},H)}
  &=& -\xi_2 |\sne_2|^2 -\xi_3 |\sne_3|^2
      -\lpa \sEC_2~\csle_2 \rpa [M_{s_2}^2] \lpa \begin{array}{c}
      \csEC_2 \\ \sle_2 \end{array}  \rpa     \\
  &-& \lpa \sEC_3~\csle_3 \rpa [M_{s_3}^2]
      \lpa \begin{array}{c} \csEC_3 \\ \sle_3 \end{array}  \rpa
      - \chi^T [M_c^2] \chi^*  - \frac{1}{2} \phi^T [M_n^2] \phi^*
\end{eqnarray*}
where $\chi^T = \lpa H_1^-~H_2^{+*}~\csne_1 \rpa $ and $\phi^T = \lpa
\varphi^T~\varphi^\dag \rpa,~ \varphi^T = \lpa H_1^0~H_2^0~\sEC_1~\sle_1 \rpa.$

For $p=2,3$ using a compact notation we have \bedm
  \xi_p = m_{\sLE_p}^2 + \frac{g^2}{2} \lpa |\sle_1|^2 - |\sEC_1|^2 \rpa
  + \frac{g^2 + g_2^2}{4} \lpa |H_1|^2 - |H_2|^2 - |\sle_1|^2 \rpa
  ~~~~~~(\Omega=2,~C=1)
\endm
On the other hand $[M_{s_2}^2] = \lpa \begin{array}{cc} Z_1 & Z_2 \\
Z_2^* & Z_3  \end{array} \rpa~(\Omega=2,~C=1)$ with
\begin{eqnarray*}
  Z_1 &=& m_{\sEC_2}^2 + \magf{Y_\mu H_1}
  + \frac{g^2}{2} \lpa |H_2|^2 + 2|\sEC_1|^2 - |\sle_1|^2
  - |H_1|^2  \rpa       \\
  Z_2 &=& Y_\mu (\mu \cH_2 + A_\mu H_1 + Y_e \csle_1 \csEC_1)  \\
  Z_3 &=& m_{\sLE_2}^2 + \magf{Y_\mu H_1}
  + \frac{g^2}{2} \lpa |\sle_1|^2 - |\sEC_1|^2 \rpa
  + \frac{g^2 - g_2^2}{4} \lpa |H_1|^2 - |H_2|^2 - |\sle_1|^2 \rpa.
\end{eqnarray*}
and $[M_{s_3}^2] = \lpa \begin{array}{cc} K_1 & K_2 \\
K_2^* & K_3  \end{array} \rpa~(\Omega=2,~C=1)$ with
\begin{eqnarray*}
  K_1 &=& m_{\sEC_3}^2 + \magf{Y_\tau H_1}
  + \frac{g^2}{2} \lpa |H_2|^2 + 2|\sEC_1|^2 - |\sle_1|^2
  - |H_1|^2  \rpa       \\
  K_2 &=& Y_\tau (\mu \cH_2 + A_\tau H_1 + Y_e \csle_1 \csEC_1)  \\
  K_3 &=& m_{\sLE_3}^2 + \magf{Y_\tau H_1}
  + \frac{g^2}{2} \lpa |\sle_1|^2 - |\sEC_1|^2 \rpa
  + \frac{g^2 - g_2^2}{4} \lpa |H_1|^2 - |H_2|^2 - |\sle_1|^2 \rpa.
\end{eqnarray*}
Similarly $[M_c^2] = \lpa \begin{array}{ccc}
    B_1      & B_2   & B_3     \\
    B_2^*    & B_4   & B_5       \\
    B_3^*    & B_5^* & B_6
\end{array} \rpa~(\Omega=2,~C=1)$ with
\begin{eqnarray*}
  B_1 &=& m_1^2 + \magf{Y_e \sEC_1}
    + \frac{g^2+g_2^2}{4} \lpa |H_1|^2 + |H_2|^2 + |\sle_1|^2 \rpa
    - \frac{g^2}{2} \lpa |\sEC_1|^2 + |H_2|^2\rpa        \\
  B_2 &=& -m_3^2 + \frac{g_2^2}{2} \cH_1 \cH_2  ~~~~~~~~~~~~~~\,
  B_3 = -Y_e(Y_e \cH_1 \csle_1 + A_e \sEC_1)
    + \frac{g_2^2}{2} \cH_1 \csle_1  \\
  B_4 &=& m_2^2 + \frac{g^2}{2} \lpa |\sEC_1|^2 - |H_1|^2\rpa
    + \frac{g^2+g_2^2}{4} \lpa |H_1|^2 + |H_2|^2 - |\sle_1|^2 \rpa    \\
  B_5 &=& \mu Y_e \sEC_1 + \frac{g_2^2}{2} H_2 \csle_1    \\
  B_6 &=&  m_{\sLE_1}^2 + \lpa |Y_e|^2 - \frac{g^2}{2} \rpa |\sEC_1|^2 +
    \frac{g^2+g_2^2}{4} \lpa |\sle_1|^2 + |H_1|^2 - |H_2|^2 \rpa
\end{eqnarray*}
and $[M_n^2] = \lpa \begin{array}{cc} S_1 & S_2^\dag \\ S_2 & S_1^*
\end{array} \rpa~(\Omega=C=1)$ with
\bedm
  S_1 = \lpa
  \begin{array}{ccccc}
      \Theta_1   & \Theta_2   & \Theta_3   & \Theta_4  \\
      \Theta_2^* & \Theta_5   & \Theta_6   & \Theta_7  \\
      \Theta_3^* & \Theta_6^* & \Theta_8   & \Theta_9  \\
      \Theta_4^* & \Theta_7^* & \Theta_9^* & \Theta_{10}
  \end{array} \rpa, \qquad ~~~\qquad  S_2 = \lpa
  \begin{array}{ccccc}
      \Lambda_1 & \Lambda_2 & \Lambda_3 & \Lambda_4  \\
      \Lambda_2 & \Lambda_5 & \Lambda_6 & \Lambda_7  \\
      \Lambda_3 & \Lambda_6 & \Lambda_8 & \Lambda_9  \\
      \Lambda_4 & \Lambda_7 & \Lambda_9 & \Lambda_{10}
  \end{array}  \rpa
\endm
\begin{eqnarray*}
  \Theta_1 &=& m_1^2 + |Y_e|^2 \lpa |\sEC_1|^2 + |\sle_1|^2 \rpa
     + \frac{g^2}{2} \lpa |\sle_1|^2 - |\sEC_1|^2 \rpa
     + \frac{g^2+g_2^2}{4} \lpa 2|H_1|^2 - |H_2|^2 - |\sle_1|^2 \rpa \\
  \Theta_2 &=& -\frac{g^2+g_2^2}{4} \cH_1 H_2   ~~~~~~~~
  \Theta_3 = \lpa |Y_e|^2 - \frac{g^2}{2} \rpa \cH_1 \sEC_1   ~~~~~~~~
  \Theta_4 = \lpa |Y_e|^2 + \frac{g^2-g_2^2}{4} \rpa \cH_1 \sle_1   \\
  \Theta_5 &=& m_2^2
     + \frac{g^2}{2} \lpa |\sEC_1|^2 - |\sle_1|^2 \rpa
     + \frac{g^2+g_2^2}{4} \lpa |\sle_1|^2 + 2|H_2|^2 - |H_1|^2 \rpa   \\
  \Theta_6 &=& \mu Y_e \csle_1 + \frac{g^2}{2} \cH_2 \sEC_1   ~~~~~~~~~~~~~~~~~
  \Theta_7 = \mu Y_e \csEC_1 - \frac{g^2-g_2^2}{4} \cH_2 \sle_1    \\
  \Theta_8 &=& m_{\sEC_1}^2 + |Y_e|^2 \lpa |H_1|^2 + |\sle_1|^2 \rpa
     + \frac{g^2}{2} \lpa 4|\sEC_1|^2 + |H_2|^2 - |H_1|^2
     - |\sle_1|^2 \rpa    \\
  \Theta_9 &=& \lpa |Y_e|^2 - \frac{g^2}{2} \rpa \sle_1 \csEC_1  \\
  \Theta_{10} &=& m_{\sLE_1}^2 + |Y_e|^2 \lpa |H_1|^2 + |\sEC_1|^2 \rpa
     + \frac{g^2}{2} \lpa 2|\sle_1|^2 - |\sEC_1|^2 \rpa
     + \frac{g^2-g_2^2}{4} \lpa |H_1|^2 - |H_2|^2 - 2|\sle_1|^2 \rpa
\end{eqnarray*}
while the corresponding elements for $S_2$ are%
\bedm
\begin{array}{llll}
  \Lambda_1 =& \displaystyle \frac{g^2+g_2^2}{4} (H_1)^2  &
  \Lambda_2 =& \displaystyle m_3^2 - \frac{g^2+g_2^2}{4} H_1 H_2  \\[3mm]
  \Lambda_3 =& \displaystyle \lpa |Y_e|^2 - \frac{g^2}{2} \rpa H_1 \sEC_1
     + Y_e A_e \csle_1 \\[3mm]
  \Lambda_4 =& \displaystyle Y_e \lpa Y_e H_1 \sle_1 + A_e \csEC_1 \rpa
     + \frac{g^2-g_2^2}{4} H_1 \sle_1
\end{array}
\endm
\bedm
\begin{array}{llll}

  \Lambda_5 =& \displaystyle \frac{g^2+g_2^2}{4} (H_2)^2 \qquad &
  \Lambda_6 =& \displaystyle \frac{g^2}{2} H_2 \sEC_1 \\[3mm]
  \Lambda_7 =& \displaystyle -\frac{g^2-g_2^2}{4} H_2 \sle_1  &
  \Lambda_8 =& \displaystyle (g \sEC_1)^2  \\[3mm]
  \Lambda_9 =&\displaystyle Y_e \lpa \mu H_2 + A_e \cH_1 \rpa +
     \lpa |Y_e|^2 - \frac{g^2}{2} \rpa \sle_1 \sEC_1 \qquad\qquad
  &\Lambda_{10} =&  \displaystyle \frac{g^2+g_2^2}{4} (\sle_1)^2
\end{array}
\endm
{\centering\section{\label{omega}Expressions for $\blm{\omega(x)}$}}
We present below expressions of $\omega(x)$ used in the definition of the
renormalization scale $Q^*$ given in Eq.~(\ref{Qstar}) for the various cases we
have considered
\\[2mm]
\textbf{\textbullet~UFB-1}
 \be
 \omega(x) = \left\{
 \begin{array}{cl}
  A_{11}(A_{12}+x)(A_{13}+x)(A_{14}+x)   & ,\quad x \leq 2.8\\
  A_{21}(A_{22}+x)(A_{23}+x)(A_{24}+x)   & ,\quad 2.8 < x \leq 3.0\\
  A_{31}(A_{32}+x)(A_{33}+x)(A_{34}+x)   & ,\quad 3.0 < x \leq 4.0\\
  A_{41}(A_{42}+x)(A_{43}+A_{44} x+x^2)  & ,\quad 4.0 < x \leq 5.0\\
  A_{51}(A_{52}+x)(A_{53}+A_{54} x+x^2)  & ,\quad 5.0 < x \leq 6.0\\
  A_{61}(A_{62}+x)(A_{63}+A_{64} x+x^2)  & ,\quad 6.0 < x \leq 7.0\\
  A_{71}(A_{72}+x)(A_{73}+A_{74} x+x^2)  & ,\quad 7.0 < x \leq 8.0\\
  A_{81}(A_{82}+x)(A_{83}+A_{84} x+x^2)  & ,\quad 8.0 < x \leq 9.0\\
  A_{91}(A_{92}+x)(A_{93}+A_{94} x+x^2)  & ,\quad 9.0 < x
 \end{array} \right.
 \ee
where
 \be
 A = \left(
 \begin{array}{cccc}
 -8.22877   &  -3.07458  &   -2.70973 &  -2.3097  \\
  1.40049   &  -3.96386  &   -3.51934 &  -2.71471  \\
  0.86922   &  -4.78838  &   -3.43647 &  -2.70523 \\
 -0.390767  &  -3.5568   &   30.1795  & -10.8231 \\
  0.0938483 &  -8.08594  &   24.5516  &  -9.49597 \\
 -0.184626  &  -4.67762  &   48.0765  & -13.5349 \\
  0.344656  &  -9.61596  &   42.2498  & -12.8772 \\
 -0.193997  &  -6.81241  &  102.729   & -19.8646 \\
  0.0313323 &  -14.349   &   63.3896  & -14.651
 \end{array} \right)
 \ee
\textbf{\textbullet~UFB-3b}
 \be
 \omega(x) = \left\{
 \begin{array}{cl}
  U_{11}(U_{12}+x)(U_{13}+x)(U_{14}+x)   & ,\quad x \leq 2.8\\
  U_{21}(U_{22}+x)(U_{23}+x)(U_{24}+x)   & ,\quad 2.8 < x \leq 3.0\\
  U_{31}(U_{32}+x)(U_{33}+x)(U_{34}+x)   & ,\quad 3.0 < x \leq 4.0\\
  U_{41}(U_{42}+x)(U_{43}+U_{44} x+x^2)  & ,\quad 4.0 < x \leq 5.0\\
  U_{51}(U_{52}+x)(U_{53}+U_{54} x+x^2)  & ,\quad 5.0 < x \leq 6.0\\
  U_{61}(U_{62}+x)(U_{63}+U_{64} x+x^2)  & ,\quad 6.0 < x \leq 7.0\\
  U_{71}(U_{72}+x)(U_{73}+U_{74} x+x^2)  & ,\quad 7.0 < x \leq 8.0\\
  U_{81}(U_{82}+x)(U_{83}+U_{84} x+x^2)  & ,\quad 8.0 < x \leq 9.0\\
  U_{91}(U_{92}+x)(U_{93}+U_{94} x+x^2)  & ,\quad 9.0 < x \\
 \end{array} \right.
 \ee
where
 \be
 U = \left(
 \begin{array}{cccc}
 -4.35862    & -3.11259 &   -2.8     &   -2.26302  \\
 -0.900195   & -3.30931 &   -2.8     &   -1.20425  \\
  0.685746   & -5.00903 &   -3.39062 &   -2.81418  \\
 -0.292473   & -3.53026 &   28.8914  &  -10.313    \\
  0.184148   & -8.25836 &   19.5607  &   -8.57881  \\
 -0.144118   & -4.62537 &   61.7438  &  -14.8604   \\
  0.192324   & -10.2975 &   37.4871  &  -11.8372   \\
 -0.125179   & -6.15239 &  114.679   &  -20.7135   \\
  0.00839312 & -18.1679 &   59.3748  &  -10.8321
 \end{array} \right)
 \ee
\textbf{\textbullet~CCB-Eb}
 \be
 \omega(x) = \left\{
 \begin{array}{cl}
  C_{11}(C_{12}+x)(C_{13}+x)(C_{14}+x)   & ,\quad x \leq 2.8\\
  C_{21}(C_{22}+x)(C_{23}+C_{24} x+x^2)   & ,\quad 2.8 < x \leq 3.0\\
  C_{31}(C_{32}+x)(C_{33}+x)(C_{34}+x)   & ,\quad 3.0 < x \leq 4.0\\
  C_{41}(C_{42}+x)(C_{43}+C_{44} x+x^2)  & ,\quad 4.0 < x \leq 5.0\\
  C_{51}(C_{52}+x)(C_{53}+C_{54} x+x^2)  & ,\quad 5.0 < x \leq 6.0\\
  C_{61}(C_{62}+x)(C_{63}+C_{64} x+x^2)  & ,\quad 6.0 < x \leq 7.0\\
  C_{71}(C_{72}+x)(C_{73}+C_{74} x+x^2)  & ,\quad 7.0 < x \leq 8.0\\
  C_{81}(C_{82}+x)(C_{83}+C_{84} x+x^2)  & ,\quad 8.0 < x \leq 9.0\\
  C_{91}(C_{92}+x)(C_{93}+C_{94} x+x^2)  & ,\quad 9.0 < x
 \end{array} \right.
 \ee
where
 \be
 C = \left(
 \begin{array}{cccc}
 -8.31888   &  -3.07077  &  -2.70937    &   -2.31018  \\
  1.9263    &  -2.71468  &  12.4319     &   -7.02268  \\
  0.581498  &  -5.27073  &  -3.4845     &   -2.68738  \\
 -0.0667458 &  -3.55197  &  53.9364     &  -13.3041   \\
 -0.0145147 &  -3.39174  & 127.123      &  -20.1434   \\
 -0.0751954 &  -3.95696  &  65.0836     &  -15.1115  \\
  0.0152963 & -19.2871   &  39.5677     &  -11.2084   \\
  0.11401   & -11.7758   &  46.3403     &  -13.0956  \\
 -0.0713367 & -6.7888    & 150.536      &  -23.613
 \end{array} \right)
 \ee
}

\end{document}